\documentclass[a4paper,11pt]{article}
\usepackage{graphicx}% Include figure files
\usepackage{bm}% bold math
\usepackage{mathptm}
\usepackage{graphics}
\usepackage{latexsym}
\usepackage{eurosym}
\usepackage{todonotes}
\usepackage{mathptmx}
\usepackage{geometry}
\usepackage{caption}
\usepackage{subcaption}
\usepackage{xcolor}
\usepackage{fullpage}
\usepackage{amssymb}
\usepackage{hyperref}
\usepackage{amsmath}
\usepackage{authblk}
\title{Transparency and enhancement in fast and slow light in $q$-deformed optomechanical system}
\author[1,2]{Akash Kundu\footnote{\href{mailto:akundu@iitis.pl}{akundu@iitis.pl}}}
\author[1]{Jaros\l aw A. Miszczak}
\affil[1]{Institute of Theoretical and Applied Informatics, Polish Academy of Sciences, Ba\l tycka 5, 44-100 Gliwice, Poland}
\affil[2]{Joint Doctoral School, Silesian University of Technology, Akademicka 2A, 44-100 Gliwice, Poland}

\date{} 

\begin{document}
\maketitle

% Author: Please give full first and last names for authors and include * after the name of all corresponding authors

% Dedication

%\dedication{Optional dedication here. If no dedication is required, please leave blank}

% Affiliations: Please provide adacemic titles (Prof. or Dr.) for all authors where applicable, and include an institutional email address for all corresponding authors

% Keywords: Please provide a minimum of three and a maximum of seven keywords, separated by commas

% Abstract should be written in the present tense and impersonal style (i.e., avoid we), and be at most 200 words long
\begin{abstract}

Nonclassical phenomena can be enhanced by introducing $q$-deformation in optomechanical systems. This motivates investigation of the optical response in a $q$-deformed linearly coupled optomechanical system. The system consists
of two deformed cavities that are linearly coupled to the motion of mechanical mirrors, and the cavities are coupled to each other by transmission strength parameter. This study shows that compared to non-deformed cases, the deformed system exhibits more rapid phase transition at the positions of
transparency windows, causing stronger and enhanced fast and slow light phenomena. Moreover, in the region $0<q<0.5$, the optomechanical system results in gain. Additionally, for a fixed deformation of cavities, by tuning the tunneling strength and optomechanical coupling, one can observe a delay and advancement in probe field in the order of milliseconds and even above milliseconds for fine-tuning of the coupling parameters. Finally, the bridge between mathematical and physical models is built by assuming the deformation to be a primitive root of unity, which indicates a class of anyon
models. These results demonstrate that $q$-deformation provides a novel method for manipulating and significantly enhancing optical phenomena not only in arbitrarily deformed optomechanical systems but also in anyon models.\\

\end{abstract}

% Text: Please use section headings and subheadings as specified below. For communications, all section headings apart from Experimental Section should be removed
% Please make the first reference to a display item bold: \textbf{Figure 1}
% Do not abbreviate Figure, Equation, etc.; display items are always singular, i.e., Figure 1 and 2.
% Equations are always singular, i.e., Equation 1 and 2, and should be inserted using the {equation} environment, not as graphics
% Please do not use footnotes in the text, additional information can be added to the Reference list.

\section{Introduction}

The concept of $q$-deformed quantum system lies in the SU$_q(2)$ algebraic group and was first introduced in \cite{sklyanin1982some,kulish1983quantum} in the context of Yang-Baxter (YB) equations which known to play a very crucial role in classical (CYB) and quantum (QYB) mechanical systems. Particularly QYB equations has a very important role in solving variety of theoretic physical problems including exactly solvable models in S-matrix theory \cite{zamolodchikov1990factorized}, 2-dimension field theories involving fields with statistics in-between bosons and fermions \cite{frohlich1992non}, conformal field theories \cite{moore1988polynomial,frenkel1988vertex,bernard1989vertex}. In Ref. \cite{macfarlane1989q} while describing the comparative structure of $q$-deformation the author quoted ``\textit{... just as the Jacobi identity is an associativity condition for a Lie algebra, so does a QYB equation play a similar role for an algebraic structure of a new type that is a generalization of a Lie algebra. The structure sometimes describes as a $q$-deformation of a Lie group.}'' Under the limit $q\rightarrow1$, the  {$q$-deformed quantum group} SU$_q(2)$ converges to the Lie algebra of SU$(2)$. 

Significant research effort has been devoted to various applications of $q$-deformation in physics. In \cite{man1995deformed} the authors presented a $q$-deformed Klein-Gordon equation with clear physical motivation to get an equation which may show nonlinear nature for higher intensities of the field. In \cite{dey2015q} an improvement in squeezing of the quadrature beyond the ordinary case has been observed using $q$-deformed parameter. In \cite{BAYINDIR2021105474} a numerical investigation of the characteristics and stabilities of the self-localized solution of nonlinear Schr\"odinger equation with a $q$-deformed Rosen-Morse potential has been conducted. Finally, very recently a $q$-deformed formalism for qubits and qutrits logic gates has been presented in \cite{altintas2014constructing,altintas2020q}, suggesting the potential  {application} of $q$-deformed systems in quantum information processing. 

In quantum optomechanics usually an optomechanical system contains information about an interacting Febry-Perot cavity \cite{PhysRevLett.98.030405} and mechanical oscillator. The impact of optomechanical systems can be seen in many fields such as, cooling of mechanical resonators \cite{arcizet2006radiationcooling,PhysRevLett.99.093901cooling,PhysRevLett.99.093902cooling,teufel2011sidebandcooling}, entanglement between macroscopic oscillator and the cavity field \cite{PhysRevA.78.032316entanglement,PhysRevA.89.014302entanglement,PhysRevLett.102.020501entanglement,PhysRevLett.88.120401entanglement}, optomechanical induced transparency (OMIT) \cite{weis2010OMIT,karuza2013OMIT,PhysRevA.90.043825OMIT}, electromagnetic induced transparency (EIT) \cite{harris1990EIT,boller1991EIT,fleischhauer2005EIT}. 
EIT  {appears} due to the effect of quantum interference occurring in three-level atoms. Observation of EIT involves two optical fields (probe and coupling). The probe field is tuned near resonance between two of the states and measures the absorption spectrum of the transition. The coupling field is much stronger and tuned near resonance at a different transition. If the states are chosen properly, the presence of coupling field will create a transparency window (TW) which will be detected by the probe. In an optomechanical system, the EIT is comparable to OMIT and can be used to tune fast and slow light \cite{thevenaz2008slow,baba2008slow,liao2020slow}. In recent years, a lot of work has been conducted to study fast and slow light
effects in a class of cavity optomechanical systems. In \cite{hussain2020fsstudy} the authors used a two atom optomechanical system to investigate fast and slow light, a comprehensive study of nonreciprocal transmission, fast and slow light effects is presented in \cite{liu2019nonreciprocalstudy}, and in \cite{wang2019mechanical} slow or fast light effect in PT-symmetric mechanical systems is conducted. A detailed study of fast and slow light effect in active-passive optomechanical cavity is studied in \cite{KUNDU2021168465}.

Motivated by the these developments, we investigate optical response in a $q$-deformed linearly coupled optomechanical system. The system consists of two deformed cavities that are linearly coupled to motion of mechanical mirrors, and the cavities are coupled to each other by transmission strength parameter. Based on the OMIT investigation, we study the fast and slow light phenomena under different deformation strength ($q$), and by manipulating various optomechanical parameters. It has been observed that under the proper selection of optomechanical parameters, the deformation strength in the system induces gain (in the range $0<q<0.5$) by arising a negative absorption peak. We also note that with the help of coupling and tunnneling parameters a clear tunability in OMIT and later strong enhancement and manipulation in super and subluminal light in deformed system can be achieved. Further when $q$ is primitive root of unity, the cavity photons behaves as anyons. In this setting we extend our investigation of phase transition and fast-slow light in case of Ising and Fibonacci anyon models. The appearance in gain and strong enhancement in super and subluminal behaviour in the probe field is also noted in case of anyonic models.

The presented manuscript is arranged  {in the following way}. In section \ref{sec:prelims} we briefly describe the preliminaries of $q$-deformation. In section \ref{sec:model} a brief description of the model Hamiltonian of the system is provided. In section \ref{sec:dynamics} we present the dynamical behavior of the system at steady state and under quantum fluctuations which is followed by a derivation of transmission amplitude.  {An investigation of OMIT, phase transition and fast-slow light phenomena is studied in section \ref{sec:results}; further compare the non-deformed to $q$-deformed cases}. In section \ref{sec:anyon} we build the bridge between the $q$-deformed mathematical formalism and physical models by considering the deformation strength a primitive root of unity which opens up a class of anyon models. Finally, in section \ref{sec:conclusion} we conclude our results.

\section{Preliminaries}\label{sec:prelims}

\subsection{$q$-deformed Harmonic Oscillator}
The algebraic application of $q$-deformation has been developed during the period $1987-89$ but the introduction of $q$-deformed harmonic oscillator \cite{macfarlane1989q,sun1989q,biedenharn1989quantum} in terms of bosonic creation and annihilation operators triggered the possibility of application of the $q$-deformation in quantum optical systems. For a normal harmonic oscillator the commutator relation satisfies the following set of equalities
\begin{eqnarray}
	&&\hat{O}\hat{O}^\dagger-\hat{O}^\dagger\hat{O}=1\;\;;\textrm{i.e.}\;\;\left[\hat{O},\hat{O}^\dagger\right]=1,\nonumber\\\label{eq:non-deformed-QHO}\\
	&&\hat{O}^\dagger\hat{O}=\hat{N}_O\;\;\;;\;\;\;\hat{O}\hat{O}^\dagger=\hat{N}_O+1,\nonumber
\end{eqnarray}
where $\hat{O}\; (\hat{O}^\dagger)$ define bosonic annihilation (creation) operator.

Let the $q$-deformed bosonic annihilation (creation) operators defined by $\hat{O}_q(\hat{O}_q^\dagger)$.  {Under $q$-deformation the equation (\ref{eq:non-deformed-QHO}) is modified as follows}
\begin{eqnarray}
	&&\hat{O}_q\hat{O}_q^\dagger-q^{\mp1}\hat{O}_q^\dagger\hat{O}_q=q^{\pm  {\hat{N}_O}},\nonumber\\\label{eq:$q$-deformed-QHO}\\
	&&\hat{O}_q^\dagger\hat{O}_q=\left[\hat{N}_O\right]\;\;\;;\;\;\;\hat{O}_q\hat{O}_q^\dagger=\left[\hat{N}_O+1\right],\nonumber
\end{eqnarray}
where $\left[\hat{N}_O\right]$ is the $q$-number representation of bosonic number operator $\hat{N}_O$ which can be  {addressed in terms of} bosonic number operator as
\begin{equation}
	\left[\hat{N}_O\right]=\dfrac{q^{ {\hat{N}_O}}-q^{- {\hat{N}_O}}}{q-q^{-1}}.\label{eq:q-number-representation}
\end{equation}
The $q$-number representation can be modified under the considerations when $q$ is a real entity and when $q$ defines phase factor as

\begin{equation}
	\left[\hat{N}_O\right] =  
	\begin{cases}
		\dfrac{\textrm{sinh}(k {\hat{N}_O})}{\textrm{sinh}(k)},      & q\rightarrow\textrm{real},\; k=\textrm{log}_eq,\\\\
		\dfrac{\textrm{sin}(k {\hat{N}_O})}{\textrm{sin}(k)}, & q\rightarrow\textrm{phase},\; k=-i\textrm{log}_eq.
	\end{cases}
\end{equation}
In the limit $q\rightarrow1$, we have
\begin{equation}
	\text{lim}_{q\rightarrow1}\left[\hat{N}_O\right]=\hat{N}_O.\label{eq:limit-to-nondeformation}
\end{equation}
A brief treatment on $q$-deformed algebra can be found in \cite{bonatsos1999quantum} where the author explicitly defined $q$-deformed Hermite \cite{floreanini1991q,van1992q,chang1992q} and Laguerre polynomials \cite{floreanini1991q}, $q$-deformed hyperbolic functions \cite{exton1983q}, and $q$-derivatives and integrals \cite{gray1990completeness,bracken1991q}.

The commutator for $q$-deformed quantum system defined by equation (\ref{eq:$q$-deformed-QHO}) can be re-written as,
\begin{equation}
	\left[\hat{O}_q,\hat{O}_q^\dagger\right]=\left[\hat{N}_O+1\right]-\left[\hat{N}_O\right] = \dfrac{q^{ {\hat{N}_O}+1}+q^{- {\hat{N}_O}}}{1+q}\label{eq:commutator-another-representation}.
\end{equation}
This equation will turn out to be extremely helpful for further work. Analogous to normal harmonic oscillator Hamiltonian
\begin{align}
	\hat{H}&=\dfrac{\hslash\omega}{2}\left[\hat{O}\hat{O}^\dagger+\hat{O}^\dagger\hat{O}\right],\nonumber\\	
	&=\dfrac{\hslash\omega}{2}\left[2\hat{N}_O+1\right],\label{eq:normal-hamiltonian}
\end{align}
the $q$-deformed Hamiltonian can written as follows
\begin{align}
	\hat{H}_q&=\dfrac{\hslash\omega}{2}\left[\hat{O}_q\hat{O}^\dagger_q+\hat{O}^\dagger_q\hat{O}_q\right]\nonumber\\	
	&=\dfrac{\hslash\omega}{2}\left([\hat{N}_O+1]+[\hat{N}_O]\right),\label{eq:$q$-deformed-hamiltonian}
\end{align}
 {which can be re-written} when $q$ is real and phase factor as follows
\begin{equation}
	\hat{H}_q =  
	\begin{cases}
		\dfrac{\hslash\omega}{2}\dfrac{\text{sinh}\left[k\left(\hat{N}_O+\dfrac{1}{2}\right)\right]}{\text{sinh}\left(\dfrac{k}{2}\right)},      & q\rightarrow\text{real},\\\\
		\dfrac{\hslash\omega}{2}\dfrac{\text{ {sin}}\left[k\left(\hat{N}_O+\dfrac{1}{2}\right)\right]}{\text{ {sin}}\left(\dfrac{k}{2}\right)}, & q\rightarrow\text{phase} {\rightarrow e^{\textrm{i}\theta}},
	\end{cases}
	\label{eq:q-hamiltonian}
\end{equation}

 \noindent{where $\theta$ is a real number. If $q$ is primitive root of unity then we can express write $\theta = \dfrac{2\pi i}{m}$ for $m=2,3,\ldots$,} in this case, the $q$-deformed annihilation and creation operator represents a discretize two-anyon radial motion frozen \cite{floratos1990quantum}.

\begin{figure}[t!]	\centering
	\begin{subfigure}{0.6\textwidth}
		\includegraphics[width=\textwidth]{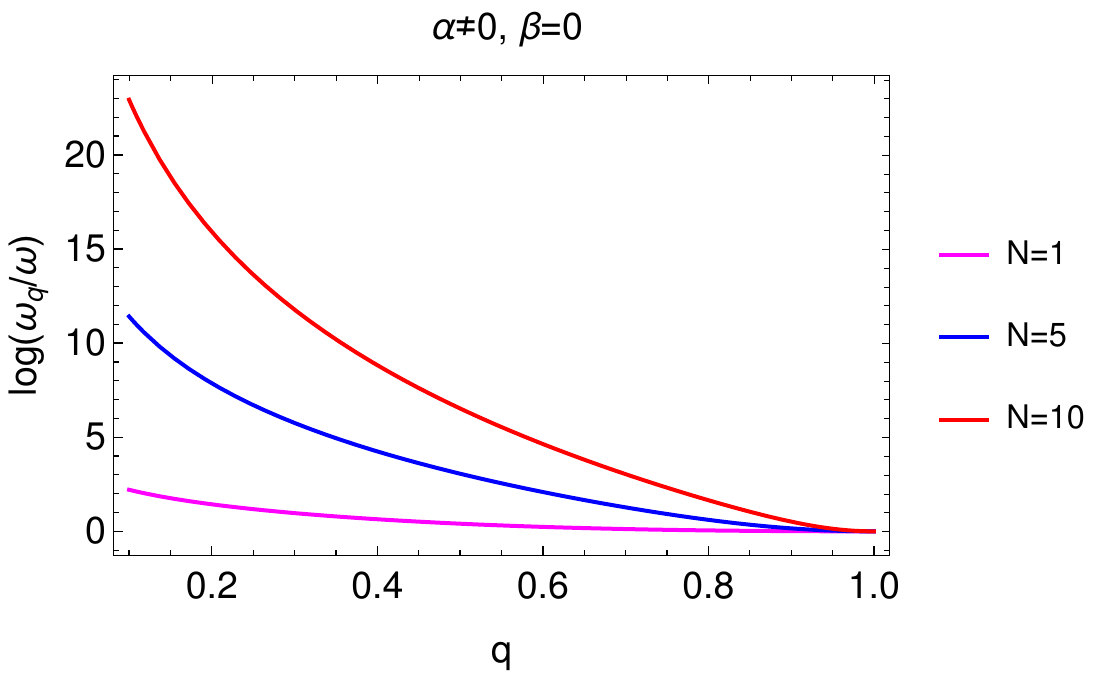}
		\caption{Real deformation parameter.}
		\label{fig.q-freq1}
	\end{subfigure}
	\begin{subfigure}{0.6\textwidth}
		\includegraphics[width=\textwidth]{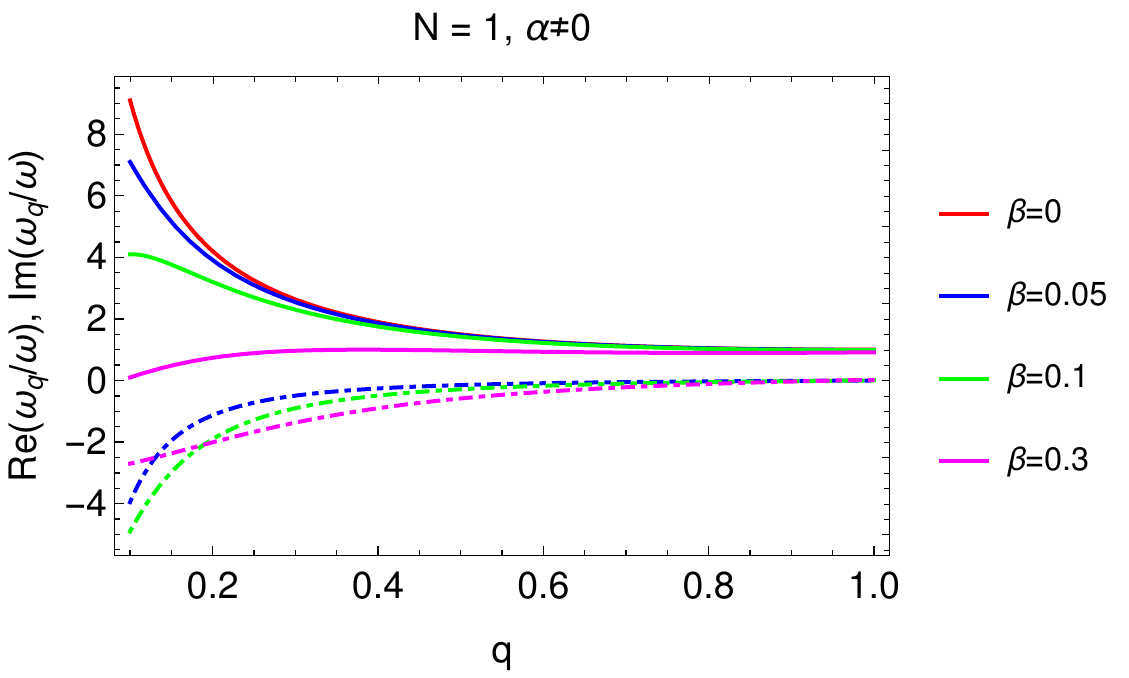}
		\caption{Complex deformation parameter.  {The solid lines are real and the dot-dashed lines are the complex part of the normalized $q$ frequency.}}
		\label{fig:q-freq2}
	\end{subfigure}
	\caption{The variation of the ratio of $q$-deformed and non-deformed frequency with $q$-parameter when (a) $q$ is real and (b) $q=\alpha+\textrm{i}\beta$.}
	\label{fig:q-freq}
\end{figure}

\subsection{Physical interpretation $q$-deformation}\label{sec:physical-interpretation-q-deform}
Many attempt were made to produce a physically sensible explanation of $q$-deformed oscillators which resulted in several publications in last few decades. Some authors connect $q$-oscillators with relativistic oscillation model \cite{chaichian1990quantum}, where other introduce $q$-oscillators and SU$_q(2)$ quantum group in the frame of generalized Jaynes-Cummins model \cite{chaichian1990quantum}. The common thing among these attempts had that they were  based purely on the mathematical properties of $q$-deformed oscillators. However, in \cite{man1993physical} while observing the classical limits of quantum q-oscillators the authors reached to a conclusion that \textit{$q$-oscillator is a normal quantum oscillator with specific non-linearity.} They also observe that this turn out to be \textit{the classical nonlinear oscillator with the frequency depending upon amplitude by particular dependence.} The statement is justified by solving the $q$-deformed Hamiltonian of harmonic oscillator
\begin{equation}
	\hat{H}_q=\hslash\omega\hat{O}^\dagger_q\hat{O}_q,\label{eq:$q$-deformed-hamiltonian-JC-form}
\end{equation}
which is $q$-deformed version of the state-of-the-art Hamiltonian
\begin{equation}
	\hat{H}=\hslash\omega\hat{O}^\dagger\hat{O}.\label{eq:non-deformed-hamiltonian-JC-form}
\end{equation}
The solution for $q$ $\rightarrow$ \textit{real} returns $q$-deformed frequency 
\begin{equation}
	\omega_q=\omega \dfrac{k\text{cosh}\left[k\left(\hat{O}^\dagger\hat{O}\right)\right]}{\text{sinh}k},\label{eq:$q$-deformed-frequency}
\end{equation}
a detailed derivation is provided in Appendix \ref{sec:q-freq}.

The conclusion one can make from the Eq.~\eqref{eq:$q$-deformed-frequency} is that the frequency of a normal harmonic oscillator is a constant entity but when it is acted by a   {deformation strength $q$}, then the frequency of the $q$-deformed harmonic oscillator depends on the deformation factor $k=\text{log}_eq$. This leads to the interpretation that, a $q$-oscillator system carries particular non-linearity. 

In normal harmonic oscillators we characterize angular frequency $\omega$ as the evolution in any of the orbits, but when the same harmonic oscillator turns into a $q$-deformed version then the angular frequency must be uniquely defined when we move from one orbit to another; which indicates that the evolution along any orbit can now be expressed as a functional constant for each orbit individually. As initially $q$ considered as real, hence the frequency of a $q$-oscillator instead of its phase depends on its amplitude. And the degree of non-linearity enhances as we increase the amplitude.

A more generalized expression of Eq.(\ref{eq:$q$-deformed-frequency}) is given by considering $q$ as a complex entity i.e. $q\rightarrow\alpha+i\beta$, which leads to the following form
\begin{equation}
	\omega_q = \omega\times\dfrac{q^{N+1}+q^{-N}}{1+q},
\end{equation}
where $N$ is the number of particles in the  {state-of-the-art} harmonic oscillator. In Fig.(\ref{fig:q-freq}), we illustrate the variation of normalized $q$-deformed frequency under arbitrary deformation strength.

\section{Model}\label{sec:model}
\begin{figure}
	\centering
	\includegraphics[width=\linewidth]{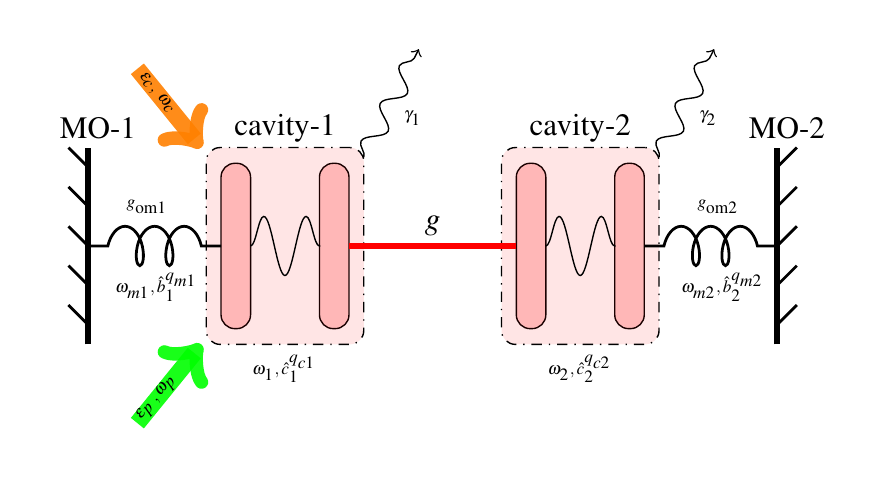}
	\caption{Illustration of the system where two interacting $q$- deformed cavities (\textbf{cavity-1} and \textbf{cavity-2}), with tunnelling strength $g$ are coupled to mechanical oscillators (\textbf{MO-1} and \textbf{MO-2}) with coupling strength $g_{\textrm{om}i}$. $\gamma_1$, and $\gamma_2$ are the loss rate of the \textbf{cavity-1} and \textbf{cavity-2} respectively.}
	\label{fig:model}
\end{figure}

Let us now describe the model used in the presented paper. As illustrated in Fig.(\ref{fig:model}), we consider an optomechanical system consisting of two deformed cavities, where each cavity's external mirrors are movable but the internal mirrors are fixed. The \textbf{cavity-1} represented by deformed annihilation (creation) operator $\hat{c}_1^{q_{\textrm{c}1}}$ ($\hat{c}_1^{q_{\textrm{c}1}\dagger}$) is coupled to the \textbf{cavity-2} represented by deformed annihilation (creation) operator $\hat{c}_2^{q_{\textrm{c}2}}$ ($\hat{c}_2^{q_{\textrm{c}2}\dagger}$) by a tunnelling parameter $g$. The  $q_{\textrm{c}1},\;q_{\textrm{c}2}$ are the deformation strength of the \textbf{cavity-1} and \textbf{cavity-2} respectively. The first cavity is coupled to the mechanical oscillator \textbf{MO-1} with deformed annihilation (creation) operator $\hat{b}^{q_{\textrm{m}1}}(\hat{b}^{q_{\textrm{m}1}\dagger})$ by coupling strength $g_{\text{om}1}$. Meanwhile the second cavity is coupled with \textbf{MO-2} with $q$-deformed annihilation (creation) operator $\hat{b}^{q_{\textrm{m}2}}(\hat{b}^{q_{\textrm{m}2}\dagger})$ by optomehcnaical coupling parameter $g_{\text{om}2}$. Where $q_{\textrm{m}1},\;q_{\textrm{m}2}$ are the deformation strength of the \textbf{MO-1} and \textbf{MO-2} respectively. The loss rate and resonance frequency of the \textbf{cavity-1} are $\gamma_1$, $\omega_1$ and for the \textbf{cavity-2} are $\gamma_2$, $\omega_2$. The decay rate and the resonance frequency of \textbf{MO-1} are $\gamma_{\text{m}1}$, $\omega_{\text{m}1}$ and for \textbf{MO-2} are $\gamma_{\text{m}2}$, $\omega_{\text{m}2}$ respectively.

In addition to the parameters motioned above the \textbf{cavity-1} is driven by a coupling field of strength $\varepsilon_c$, frequency $\omega_c$ and a weak probe field $\varepsilon_p$ of frequency $\omega_p$. The total Hamiltonian of the cavity optomechanical system in the rotating frame at frequency $\omega_c$ is given as follows:
\begin{multline}
	\hat{H}=\sum_{i=1}^{2}\Delta_i \hat{c}^{q_{\textrm{c}i}\dagger }_i\hat{c}_i^{q_{\textrm{c}i}} + \sum_{i=1}^{2}\omega_{\textrm{m}i}\hat{b}^{q_{\textrm{m}i}\dagger}_i\hat{b}_i^{q_{\textrm{m}i}} -\sum_{i=1}^{2}g_{\textrm{om}i}\hat{c}^{q_{\textrm{c}i}\dagger }_i\hat{c}_i^{q_{\textrm{c}i}}\left(\hat{b}^{q_{\textrm{m}i}\dagger }_i
	+\hat{b}_i^{q_{\textrm{m}i}}\right) +\textrm{i}\varepsilon_c\left(\hat{c}^{q_{\textrm{c}1}\dagger }_1 - \hat{c}_1^{q_{\textrm{c}1}} \right)\\
	+ \textrm{i}\varepsilon_p \left(\hat{c}^{q_{\textrm{c}1}\dagger}_1 e^{-\textrm{i}\Delta t} - \hat{c}_1^{q_{\textrm{c}1}}e^{\textrm{i}\Delta t} \right) +g\left(\hat{c}^{q_{\textrm{c}2}\dagger}_2 \hat{c}^{q_{\textrm{c}1}}_1 + \hat{c}^{q_{\textrm{c}1}\dagger}_1 \hat{c}^{q_{\textrm{c}2}}_2\right). \label{eq:deformed-hamiltonian}
\end{multline}
In Eq.(\ref{eq:deformed-hamiltonian}) the first two terms, denote the free energy of the two cavities, and the movable mirrors respectively, with $\Delta_i = \omega_{i} - \omega_c$ denoting the detuning of the cavities (for $i=1,
\;2$), and $\omega_{\text{m}i}$ are the oscillator frequencies. The third term represents the linear coupling between the cavities and the respective mechanical oscillators. The fourth and the fifth terms correspond to the interaction of the \textbf{cavity-1} with the coupling field of strength $\varepsilon_{c}$, and the probing field of strength $\varepsilon_{p}$ respectively, where $\Delta = \omega_p-\omega_c$. The last term describes the hopping between the two cavities with hopping strength $g$. For linearly coupled system the optomechanical coupling strength defined as  
\begin{equation}
	g_{\text{om}i} = \dfrac{\delta\omega_{i}}{\delta Q_{i}^{q_{\textrm{m}i}}}\hat{Q}^2_{\textrm{ZPF, i}},\;\text{where}\;\;\hat{Q}_{\textrm{ZPF}, i} = \sqrt{\dfrac{1}{m\omega_{\text{m}i}}},
\end{equation}
is the zero-point fluctuations of the mechanical membrane's displacement and $m$ is the effective mass. The quadratures of deformed system $\hat{Q}_{i}^{q_{\textrm{m}i}} = \left(\hat{b}_{i}^{q_{\textrm{m}i}} + \hat{b}_{i}^{q_{\textrm{m}i}\dagger}\right)/\sqrt{2}$, and $\hat{P}_{i}^{q_{\textrm{m}i}} = \left(\hat{b}_{i}^{q_{\textrm{m}i}} - \hat{b}_{i}^{q_{\textrm{m}i}\dagger}\right)/\sqrt{2}\textrm{i}$.

\section{Dynamics of the System}\label{sec:dynamics}
To get useful information about the system we are going to observe the dynamics of the system by using \textit{quantum Langevin equation}, which returns us a set of dynamical equations as follows:
%\begin{widetext}
\begin{align}
	& {\frac{d\hat{c}_1^{q_{\textrm{c}1}}}{dt}}  =\left(-i\Delta_1\hat{c}_1^{q_{\textrm{c}1}} +\varepsilon_c + \varepsilon_p e^{-i\Delta t} + ig_{\textrm{om}1}\hat{c}_1^{q_{\textrm{c}1}} \left(\hat{b}_1^{q_{\textrm{m}1}\dagger} + \hat{b}_1^{q_{\textrm{m}1}}\right)-ig\hat{c}_2^{q_{\textrm{c}2}}\right) \left[\hat{c}_1^{q_{\textrm{c}1}}, \hat{c}_1^{q_{\textrm{c}1}\dagger}\right]-\kappa_1\hat{c}_1^{q_{\textrm{c}1}}+\sqrt{\kappa_1}\hat{c}_{1,\text{in}}^{q_{\textrm{c}1}},\nonumber\\
	& {\frac{d\hat{c}_2^{q_{\textrm{c}2}}}{dt}} = \left(-i\Delta_2\hat{c}_2^{q_{\textrm{c}2}} + ig_{\textrm{om}2} \hat{c}_2^{q_{\textrm{c}2}} \left(\hat{b}_2^{q_{\textrm{m}2}\dagger} + \hat{b}_2^{q_{\textrm{m}2}}\right) -ig\hat{c}_1^{q_{\textrm{c}1}}\right) \left[\hat{c}_2^{q_{\textrm{c}2}}, \hat{c}_2^{q_{\textrm{c}2}\dagger}\right] - \kappa_2\hat{c}_2^{q_{\textrm{c}2}} +\sqrt{\kappa_2}\hat{c}_{2,\text{in}}^{q_{\textrm{c}2}}, \nonumber \\
	& {\frac{d\hat{b}_1^{q_{\textrm{m}1}}}{dt}} = \left(-i\omega_{\textrm{m}1}\hat{b}_1^{q_{\textrm{m}1}} + ig_{\textrm{om}1}\hat{c}_1^{q_{\textrm{c}1}\dagger}\hat{c}_1^{q_{\textrm{c}1}}\right) \left[\hat{b}_1^{q_{\textrm{m}1}}, \hat{b}_1^{q_{\textrm{m}1}\dagger}\right] - \gamma_{\textrm{m}1}\hat{b}_1^{q_{\textrm{m}1}}+\sqrt{\gamma_{\text{m}1}}\hat{b}_{1,\text{in}}^{q_{\textrm{m}1}}, \nonumber\\
	& {\frac{d\hat{b}_2^{q_{\textrm{m}2}}}{dt}} = \left(-i\omega_{\textrm{m}2}\hat{b}_2^{q_{\textrm{m}2}} + ig_{\textrm{om}2}\hat{c}_2^{q_{\textrm{c}2}\dagger}\hat{c}_2^{q_{\textrm{c}2}}\right) \left[\hat{b}_2^{ {q_{\textrm{m}2}}}, \hat{b}_2^{q_{\textrm{m}2}\dagger}\right] - \gamma_{\textrm{m}2}\hat{b}_2^{q_{\textrm{m}2}}+\sqrt{\gamma_{\text{m}2}}\hat{b}_{2,\text{in}}^{q_{\textrm{m}2}}.\nonumber
\end{align}
After simplification we get,
\begin{subequations}
	\begin{align}
		& {\frac{d\langle \hat{c}_1^{q_{\textrm{c}1}} \rangle}{dt}} = \left[-\left(\kappa_1^{q_{\textrm{c}1}} + i\Delta_1\right)\langle \hat{c}_1^{q_{\textrm{c}1}} \rangle + ig_{\textrm{om}1} \langle \hat{c}_1^{q_{\textrm{c}1}} \rangle \left( \langle \hat{b}_1^{q_{\textrm{m}1}\dagger} \rangle + \langle \hat{b}_1^{q_{\textrm{m}1}} \rangle\right) - ig \langle \hat{c}_2^{q_{\textrm{c}2}} \rangle + \varepsilon_c + \varepsilon_pe^{-\Delta t}\right]\chi_{N_{\textrm{c}1}}^{q_{\textrm{c}1}}\label{eq-ave1},\\
		& {\frac{d\langle \hat{c}_2^{q_{\textrm{c}2}} \rangle}{dt}} = \left[-\left(\kappa_2^{q_{\textrm{c}2}} + i\Delta_2\right)\langle \hat{c}_2^{q_{\textrm{c}2}} \rangle + ig_{\textrm{om}2} \langle \hat{c}_2^{q_{\textrm{c}2}} \rangle \left( \langle \hat{b}_2^{q_{\textrm{m}2}\dagger} \rangle + \langle \hat{b}_2^{q_{\textrm{m}2}} \rangle\right) - ig \langle \hat{c}_1^{q_{\textrm{c}2}} \rangle\right]\chi_{N_{\textrm{c}2}}^{q_{\textrm{c}2}},\\
		& {\frac{d\langle \hat{b}_1^{q_{\textrm{m}1}} \rangle}{dt}} = \left[ -\left( \gamma_{\textrm{m}1}^{q_{\textrm{m}1}} + i\omega_{\textrm{m}1} \right)\hat{b}_1^{q_{\textrm{m}1}} +ig_{\textrm{om}1} \lvert\langle \hat{c_1^{q_{\textrm{c}1}}} \rangle \rvert^2  \right]\chi_{N_{\textrm{m}1}}^{q_{\textrm{m}1}}, \\
		& {\frac{d\langle \hat{b}_2^{q_{\textrm{m}2}} \rangle}{dt}} = \left[ -\left( \gamma_{\textrm{m}2}^{q_{\textrm{m}2}} + i\omega_{\textrm{m}2} \right)\hat{b}_2^{q_{\textrm{m}2}} +ig_{\textrm{om}2} \lvert\langle \hat{c}_2^{q_{\textrm{c}2}} \rangle \rvert^2  \right]\chi_{N_{\textrm{m}2}}^{q_{\textrm{m}2}}\label{eq-ave2}.,
	\end{align}
\end{subequations}
%\end{widetext}
where $\chi_{N_\textrm{o}}^{q_\textrm{o}} = \dfrac{q_\textrm{o}^{N_\textrm{o}+1}+q_\textrm{o}^{-N_\textrm{o}}}{1+q_\textrm{o}}$, is the deformation coefficient for the optomechanical system and the substript $``\textrm{o}"$ takes values $``\textrm{c}1",\; ``\textrm{c}2"$ for the first and second cavities respectively and $``\text{m}1",\;``\text{m}2"$ for the mechanical oscillators respectively. Meanwhile $\gamma_{\textrm{m}i}^{q_{\textrm{m}i}} = \gamma_{\textrm{m}i} / \chi_{N_{\text{m}i}}^{q_{\textrm{m}i}} $,  $\kappa_i^{q_{\textrm{c}i}} = \kappa_i / \chi_{N_{\textrm{c}i}}^{q_{\textrm{c}i}} $. Also $\hat{c}_{i,\text{in}}^{q_{\textrm{c}i}}, \hat{b}_{i,\text{in}}^{q_{\textrm{c}i}}$ are the input vacuum noise of the cavity and mechanical oscillator that satisfies the following relationships:
$\langle\hat{c}_{i,\text{in}}^{q_{\textrm{c}i}}\rangle = \langle\hat{b}_{i,\text{in}}^{q_{\textrm{m}i}}\rangle = 0$, due to the fact that $\langle\hat{c}_{i,\text{in}}\rangle = \langle\hat{b}_{i,\text{in}}\rangle = 0$.
\subsection{Dynamics under fluctuations}
To observe the dynamics of the system we decompose the expectation value of the observable as
\begin{equation}
	\langle \hat{O}(t) \rangle = \bar{O} + \delta\hat{O}(t)\;,\; \hat{O} = \hat{c}_i^{q_{\textrm{c}i}},\;\hat{b}_i^{q_{\textrm{m}i}};\;\;\;\; i=1,2\label{eq: dynamics-eqn},
\end{equation}
where $\bar{O}$ corresponds to the steady state solution in the absence of probing field i.e. $\omega_p = 0$ and $\delta O(t) = \hat{O}_{-}e^{-i\Delta t} + \hat{O}_{+}e^{i\Delta t}$, induces by the weak probing field. By using the dynamical equation we can find out the steady state solutions of the modes of cavity and mechanical oscillators. At the same time an array of equations can be found out by expanding corresponding to $\delta\hat{O}(t)$, which will eventually leads us to the solutions for $\hat{O}_{\pm}$. However, in this paper we only focus on the the expression of $\hat{O}_{-}$ as we will be focusing only the resolved sidebands regime i.e. $\omega_{\text{m}i}\gg\kappa_i$, and $\Delta_i = \tilde{\Delta}_i = \omega_{\text{m}i}$ where $\tilde{\Delta}_i = \Delta_i-g\bar{\alpha}_i$,  {where $\bar{\alpha}_i$ is defined in Appendix \ref{sec:method} and defined by 
\begin{equation}	
	\bar{\alpha}_i = \dfrac{2\omega_{\textrm{m}i}g_{\textrm{om}i}}{\left(\gamma_{\textrm{m}i}^{{q_{\textrm{m}i}}}\right)^2 + \omega_{\textrm{m}i}^{^2}}\lvert \bar{c}_i^{q_{\textrm{c}i}} \rvert^2.
\end{equation}}
\subsection{Output field}
In order to reveal the OMIT, we have to study the response of the system to probing frequency, which can be detected by the output field. Based on the input–output theory, we obtain the relationship
\begin{equation}
	\varepsilon_{\textrm{out}, p}e^{-i\Delta t} + \varepsilon_{ p}e^{-i\Delta t} + \varepsilon_c = 2\kappa_1\langle c_1\rangle.
\end{equation}
If we put Eq.~(\ref{eq: dynamics-eqn}) into the above equation we get
\begin{equation}
	\mathcal{A}_T = \dfrac{\varepsilon_{\textrm{out}, p}}{\varepsilon_{ p}} + 1 = \dfrac{2\kappa_1\hat{c}_{1,-}}{\varepsilon_{ p}},\label{eq:A_T}
\end{equation}
\noindent{where $\mathcal{A}_T$ is the output probing field.} The $q$-deformed version of $\hat{c}_{1,-}$ is give as follows:
\begin{equation}
	c_{1,-}^{q_{\textrm{c}1}} = \dfrac{\varepsilon_p}{\left(\kappa_1^{q_{\textrm{c}1}} - i \beta_{N_{\textrm{c}1}}^{q_{\textrm{c}1}}\right) + \dfrac{\lvert G_1^{q_{\textrm{c}1}}\rvert^2}{\gamma_{\textrm{m}1}^{q_{\textrm{m}1}} - i \beta_{N_{\textrm{m}1}}^{q_{\textrm{m}1}}} + \dfrac{g^{2}}{\left(\kappa_2^{q_{\textrm{c}2}} - i \beta_{N_{c2}}^{q_{\textrm{c}2}}\right) + \dfrac{\lvert G_2^{q_{\textrm{c}2}}\rvert^2}{\gamma_{\textrm{m}2}^{q_{\textrm{m}2}} - i \beta_{N_{\textrm{m}2}}^{q_{\textrm{m}2}}} }},
\end{equation}
which can decomposed into real an imaginary part. Further, we get the deformed $\mathcal{A}^T$ using the deformed version of Eq.(\ref{eq:A_T}) as follows:
\begin{eqnarray}
	\mathcal{A}_T^{q_{\textrm{c}1}} &=& \dfrac{2\kappa_1^{q_{\textrm{c}1}}}{\left(\kappa_1^{q_{\textrm{c}1}} - i \beta_{N_{\textrm{c}1}}^{q_{\textrm{c}1}}\right) + \dfrac{\lvert G_1^{q_{\textrm{c}1}}\rvert^2}{\gamma_{\textrm{m}1}^{q_{\textrm{m}1}} - i \beta_{N_{\textrm{m}1}}^{q_{\textrm{m}1}}} + \dfrac{g^{2}}{\left(\kappa_2^{q_{\textrm{c}2}} - i \beta_{N_{c2}}^{q_{\textrm{c}2}}\right) + \dfrac{\lvert G_2^{q_{\textrm{c}2}}\rvert^2}{\gamma_{\textrm{m}2}^{q_{\textrm{m}2}} - i \beta_{N_{\textrm{m}2}}^{q_{\textrm{m}2}}} }},\nonumber\\ 
	&=& \mu^{q_{\textrm{c}1}}+i\delta^{q_{\textrm{c}1}},\label{eq:transmission-amp}
\end{eqnarray}
 {where $\mathcal{A}_T^{q_{\textrm{c}1}}$ is the deformed output probing field. The $\mu^{q_{\textrm{c}1}}$, and $\delta^{q_{\textrm{c}1}}$ are the deformed in- and out-of phase quadratures of the deformed probing field. For an elaborated derivation of Eq.(\ref{eq:transmission-amp}) see Appendix~\ref{sec:method}.}

\section{Results}\label{sec:results}
We now demonstrate the behaviour of the studied system. Based on previous studies on $N$ cavities which are connected through tunnelling parameter, the properties of the output field for the considered system is investigated using the parameters \cite{lin2010coherent,zheng2012controllable,sohail2016optomechanically} as follows.
\begin{itemize}
	\item $\omega_{\textrm{m}1} = \omega_{\textrm{m}2} = 2\pi\times51.8$ MHz,
	\item $\gamma_{\textrm{m}1} = \gamma_{\textrm{m}2} = 2\pi\times41$ KHz,
	\item $\kappa_1 = 2\pi\times15$ MHz,
	\item $\kappa_2 = 2\pi\times0.027$ MHz,
	\item $g = \kappa_1$,
	\item $G_1 = G_2 = 2\pi\times10$ MHz,	
\end{itemize}
if not stated otherwise. We assume that $N_{\textrm{m}1} = N_{\textrm{m}2} = N_{\textrm{c}1} = N_{\textrm{c}2} = 1$.

\subsection{Non-deformed case}
Let us start by focusing in the case of the considered system when $q_{\textrm{c}1}=q_{\textrm{c}2}=q_{\textrm{m}1}=q_{\textrm{m}2}=1$. All the remaining optomechanical parameters are same as mentioned above.

\subsubsection{Optomechanical induced transparency }
\begin{figure}[b!]
	\centering
	\includegraphics[width=\linewidth]{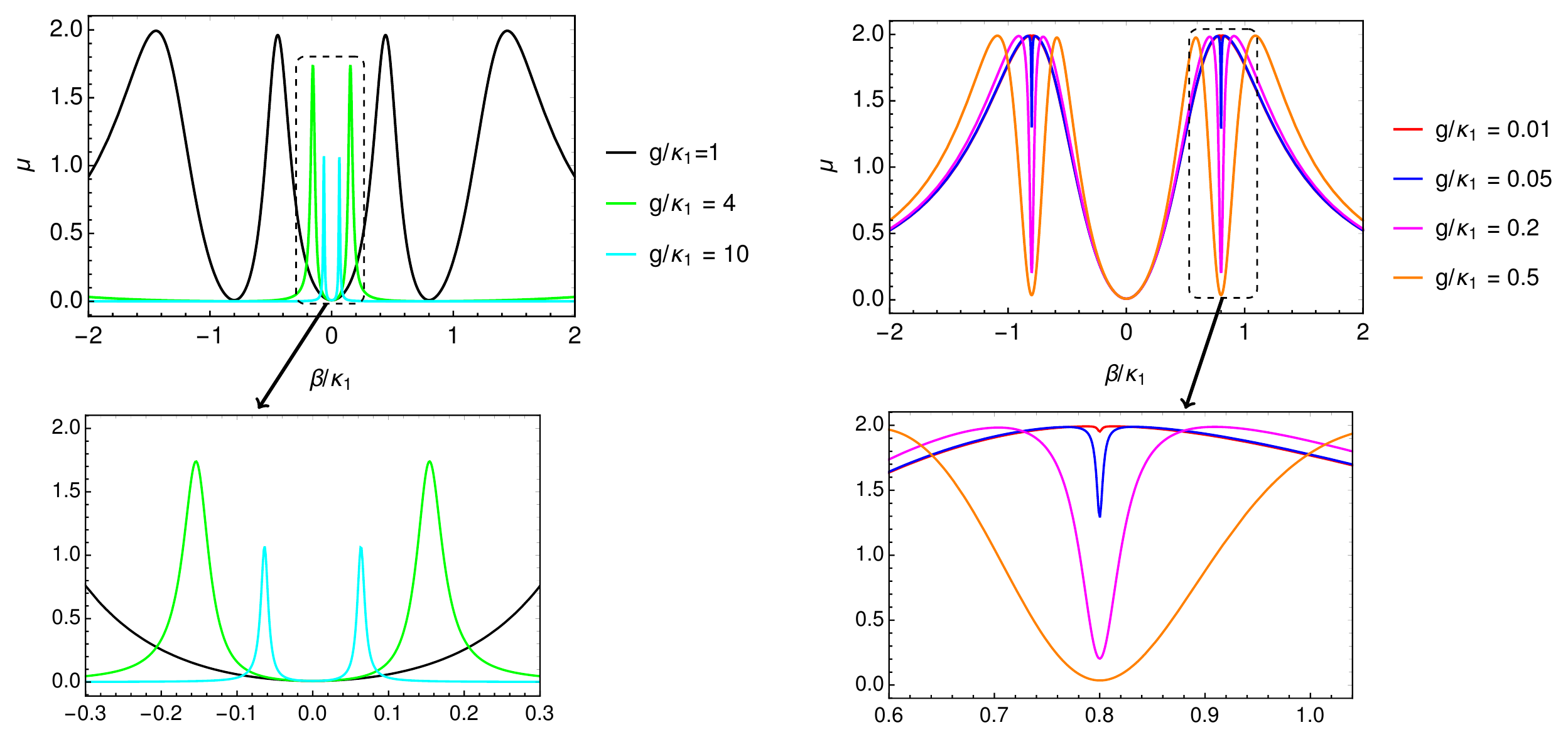}
	\caption{Illustration of variation of the $\mu$ with $\beta/\kappa_1$ when $G_1=G_2$ and $q=1$.}
	\label{fig:q-transmission-non-deformed}
\end{figure}

In the non-deformed case we observe how the tunnelling parameter effects the appearance of the transparency windows (TW). As illustrated in Fig.(\ref{fig:q-transmission-non-deformed}), both the optomechanical couplings i.e. $G_1$ and $G_2$ are non-zero in the system which causes appearance of an additional TW at $\beta/\kappa_1\approx0$ where $\beta = \Delta-\omega_m$, along with the TWs appear at $\beta/\kappa_1 \approx \pm 0.80$. It should be noted that when $G_1=0$ the central TW vanishes~\cite{sohail2016optomechanically}.

For $g >\kappa_1$, as we increase the tunnelling parameter we observe the TW at $\beta/\kappa_1 \approx 0$ becomes narrower and the adjacent absorption peaks turns smaller. Finally for  {$g/\kappa_1>4$,} the absorption peaks that was previously appearing both side of the $\beta/\kappa_1<1$, diminishes finally we get a single wide TW. This is illustrated in the left picture of Fig.(\ref{fig:q-transmission-non-deformed}).

Meanwhile when $g<\kappa_1$, and varrying $g$ in range $0\le g<1$, helps us observe that the TW that appeared at the position $\beta/\kappa_1 = \pm0.8$, turns narrower and shallower. And at $g=0$, the TWs at $\beta/\kappa_1$ vanishes, leaving only the central TW. This is illustrated in the right hand side picture of Fig.(\ref{fig:q-transmission-non-deformed}).

\subsubsection{Fast-slow light}
The fast and slow light effects in the weak probe field are illustrated and discussed in this section. For an optomechanical system, in the region of narrow TWs a rapid phase transition occurs using the relation
\begin{equation}
	\phi(\omega_p) = \textrm{arg}[A_T(\omega_p)],
\end{equation}
this rapid phase transition can cause the group delay in output field. In respect with input-output relationship \cite{aspelmeyer2014cavity}, the optical group delay of transmitted light is related to $\phi$ as follows
\begin{equation}
	\tau = \dfrac{\partial\phi(\omega_p)}{\partial\omega_p} = \dfrac{\partial\phi(\omega_p)}{\partial\beta},
	\label{eq:group-delay}
\end{equation}
where $\beta_i = \Delta-\omega_{mi}$. In Eq.(\ref{eq:group-delay}), $\tau>0$ indicates \textit{slow-light}, and $\tau<0$ denotes \textit{fast-light}.
\begin{figure}[tbh!]
	\centering
	\includegraphics[width=\linewidth]{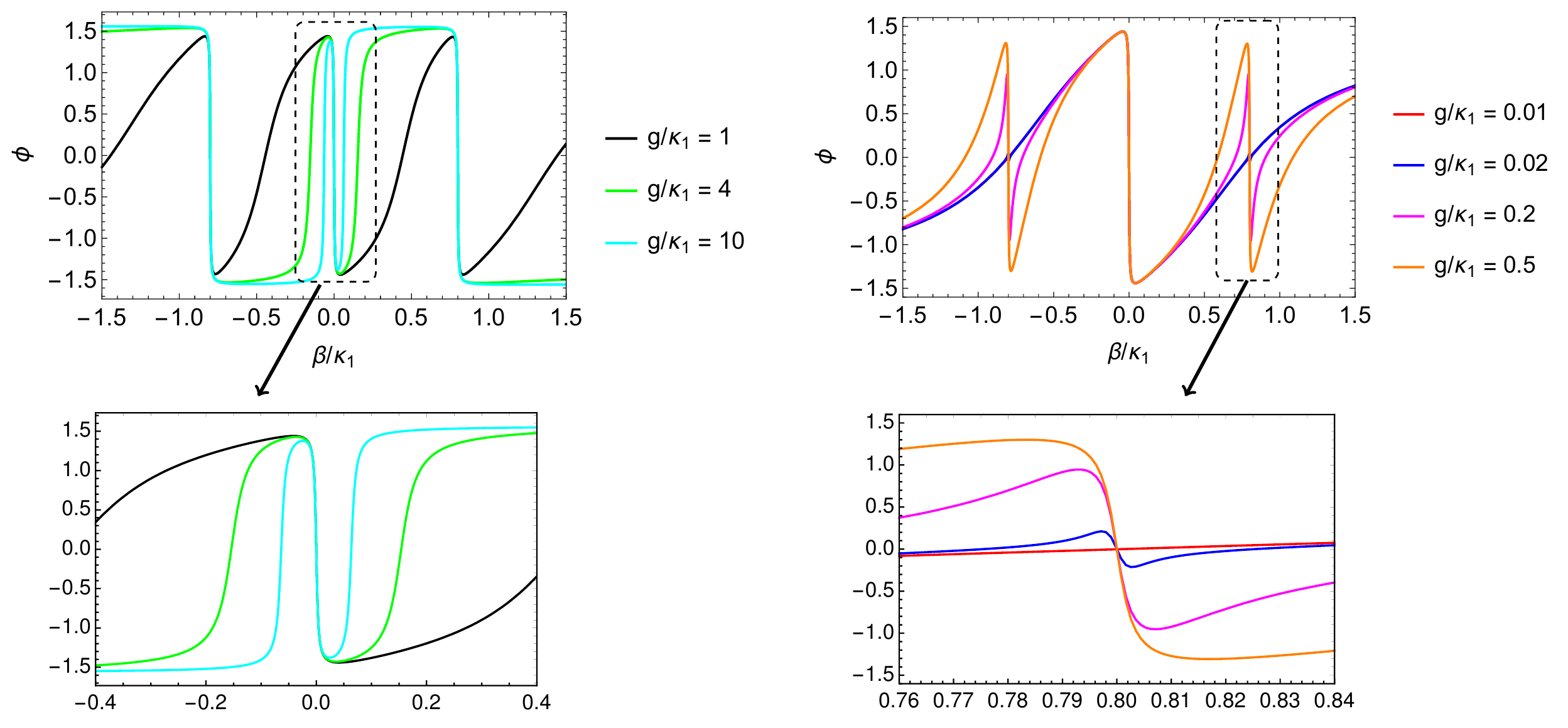}
	\caption{The phase of the probe field as a function of $\beta/\kappa_1$.}
	\label{fig.phase_non-deformed_q1_G1eqG2}
\end{figure}

\begin{figure}[tbh!]
	\centering
	\includegraphics[width=\textwidth]{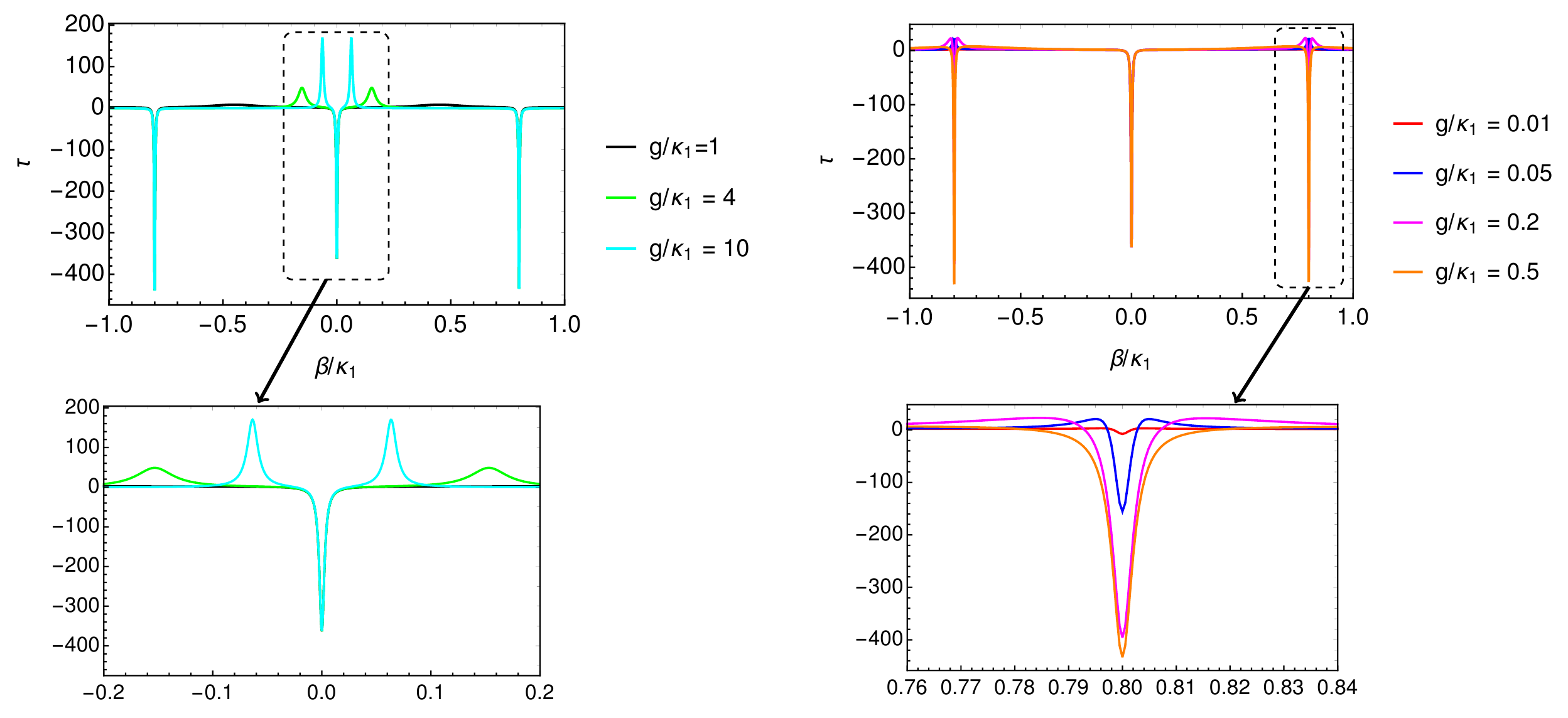}
	\caption{The group delay of the probe field as a function of $\beta/\kappa_1$.}
	\label{fig.delay_non-deformed_q1_G1eqG2}
\end{figure}

Through Fig.(\ref{fig.phase_non-deformed_q1_G1eqG2}) it can be observed that the rapid phase transition occurs at $\beta/\kappa_1\approx\pm0.8$ and $\beta/\kappa_1 = 0$, which are indeed the positions where the transparency windows appear. Interestingly it has been observed that the manipulation of phase transition near the origin is possible for $g>\kappa_1$, illustrated in the left hand side picture of Fig.(\ref{fig.phase_non-deformed_q1_G1eqG2}). On the other hand phase transition appears at $\beta/\kappa_1\approx\pm0.8$ can be manipulated by varying the tunnelling parameter in the range $0 < g < \kappa_1$, and at $g=0$, the phase transition in these regions vanishes. The appearance of sharp enhancement and decrement in phase of the probe field indicated that at that particular normalized detuning group delay can be strongly manipulated. We plot the group delay, $\tau$ (defined in Eq.(\ref{eq:group-delay})) with the normalized detuning $\beta/\kappa_1$ as illustrated in the sub-figures of Fig.(\ref{fig.delay_non-deformed_q1_G1eqG2}).
It is observed that when $g<\kappa_1$, strongest superluminal (fast light) phenomena appears at the $\beta/\kappa_1\approx\pm0.8$ (with pulse advancement of the order of 450 $\mu s$) and $\beta/\kappa_1=0$ (pulse advancement by $\approx$ 380 $\mu s$), but subluminal (slow light) phenomena is weak. By varying the tunnelling parameter in range $0<g<1$ we can manipulate the strength of fast light as shown on the right hand side of Fig.(\ref{fig.delay_non-deformed_q1_G1eqG2}). However, it does not put any impact on the central superluminal peak.

On the other hand, by varying $g>\kappa_1$, we can induce subluminal phenomena, which can be seen though two symmetric peaks appears across the central superluminal peak illustrated on the left hand side picture of Fig.(\ref{fig.delay_non-deformed_q1_G1eqG2}). Interestingly the strength of the sublimunal peak completely depends upon the strength of the tunnelling between two non-deformed cavities. As the tunnelling between cavities becomes stronger the subluminal peaks moves more and more closer to $\beta/\kappa_1=0$, simultaneously the superluminal peak deminishes. Finally, the two symmetric subluminal peaks gets merged and exists as a single peak, which causes the pulse delay of the order of 380 $\mu s$ at $g = 5\times10^3\kappa_1$.

\begin{figure}
	\centering
	\includegraphics[width=0.7\textwidth]{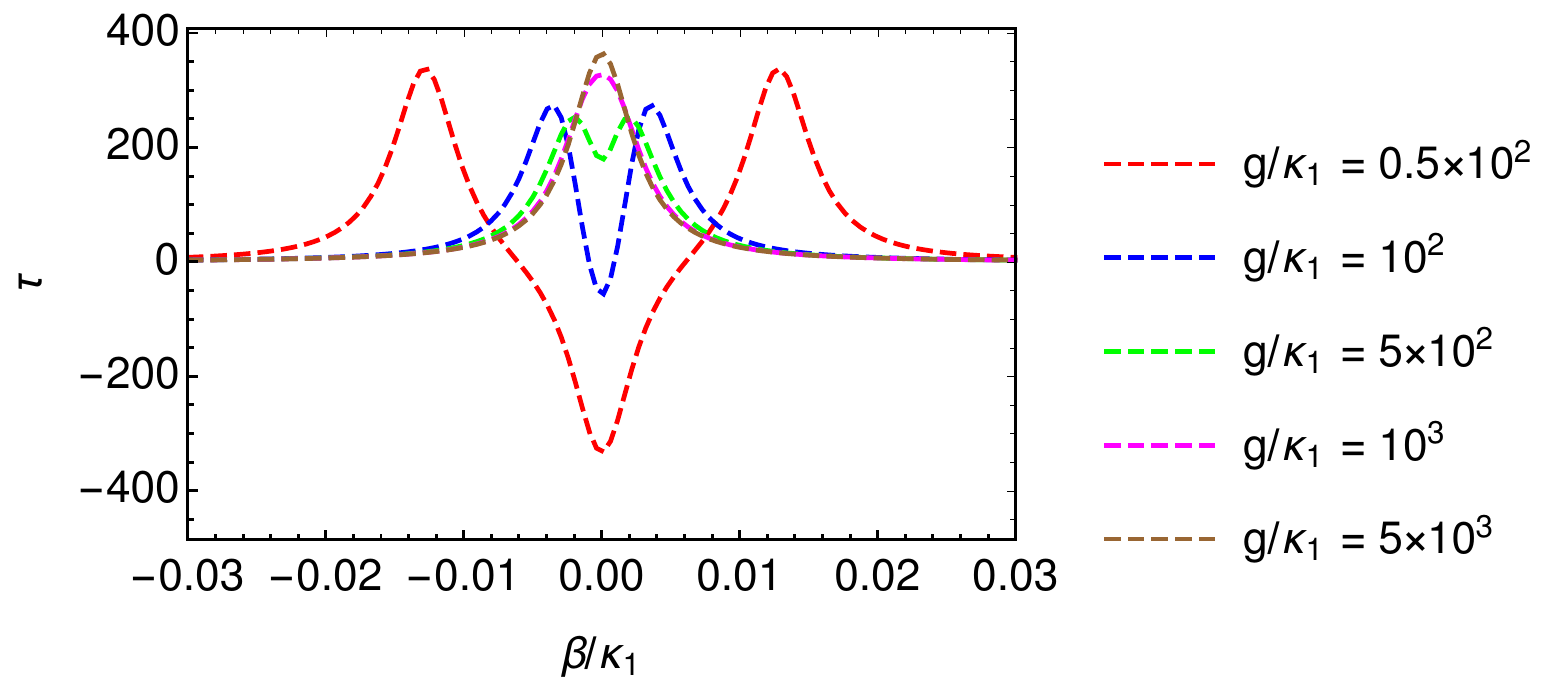}
	\caption{Group delay around $\beta/\kappa_1=0$ for strong inter-cavity tunnelling.}
\end{figure}

Experimentally the value of tunnelling strength is considered to be strong when $g>(\kappa_1+\kappa_2)/2$, and otherwise weak \cite{liao2020slow}. In our setup if $g>7.5135\kappa_1$, we can define the inter-cavity transmission to be strong. In this range even though we observe a strong superluminal phenomena at the positions of the TWs but the subluminal phenomena is smaller than the half compared to superluminal maxima. Our aim is to achieve strong super and subluminal effect at experimentally achievable tunnelling strength. We, in the following section show that it can be achieved by introducing a slight deformation to the cavities.

\subsection{$q$-deformed case}
\begin{figure}[b!]
	\centering
	\includegraphics[width=0.7\linewidth]{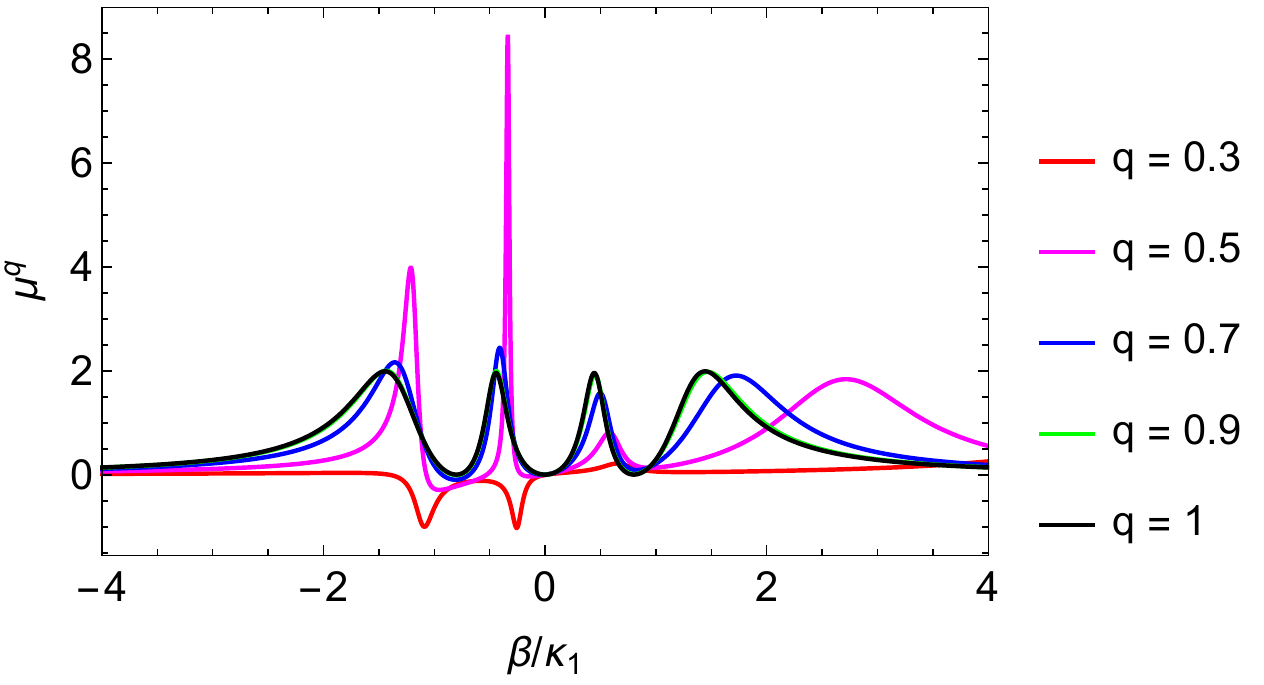}
	\caption{Real component of $\mathcal{A}_T^q$ for different deformation amplitude of the cavities for $G_1=G_2$.}
	\label{fig:q-omit}
\end{figure}

In this section we present the nontrivial nature and corresponding advantages in the optomechanical system by introducing nonlinearity in the system by tuning the deformation strength in the cavities. We compare and observe the deformed OMIT, phase and fast-slow light with the nondeformed investigation. Before discussing the results and advantages of introducing deformation we formulate the deformed version of the cavity parameters in terms of the non-deformed ones as follows
\begin{eqnarray}
	&&\dfrac{G_1^{q_{\textrm{c}1}}}{G_1} = \dfrac{\bar{c}_1^{q_{\textrm{c}1}}}{c_1} = \dfrac{\kappa_1+i\omega_{\textrm{m}1}+\dfrac{g^2}{\kappa_2+i\omega_{\textrm{m}2}}}{\kappa_1^{q_{\textrm{c}1}}+i\omega_{\textrm{m}1}+\dfrac{g^2}{\kappa_2^{q_{\textrm{c}2}}+i\omega_{\textrm{m}2}}},\label{eq:$q$-deformed-params1}\\
	&&\dfrac{G_2^{q_{\textrm{c}2}}}{G_2} = \left(\dfrac{\kappa_2+i\omega_{\textrm{m}2}}{\kappa_2^{q_{\textrm{c}2}}+i\omega_{\textrm{m}2}}\right)\times\dfrac{\bar{c}_1^{q_{\textrm{c}1}}}{c_1},\label{eq:$q$-deformed-params2}\\
	&&\gamma_{\textrm{m}i}^{q_{\textrm{m}i}} = \gamma_{\textrm{m}i}\times\left(\chi_{N_{\textrm{m}i}}^{q_{\textrm{m}i}}\right)^{-1};\;\; \kappa_{i}^{q_{\textrm{c}i}} = \kappa_{i}\times\left(\chi_{N_{\textrm{c}i}}^{q_{\textrm{c}i}}\right)^{-1},
	\label{eq:$q$-deformed-params3}
\end{eqnarray}
where $\bar{c}_i$ represents the steady state value of $\hat{c}_i$, and recalling $\chi_{N_\textrm{o}}^{q_\textrm{o}} = \dfrac{q_\textrm{o}^{N_\textrm{o}+1}+q_\textrm{o}^{-N_\textrm{o}}}{1+q_\textrm{o}}$, is the deformation coefficient for the optomechanical system and the substript $``\textrm{o}"$ takes values $``\textrm{c}1",\; ``\textrm{c}2",$ the cavity modes and $``\text{m}1",\;``\text{m}2"$ for the mechanical modes. Throughout the calculations we have used the the same optomechanical parameter values as used in the case of non-deformed investigation. Meanwhile one can introduce deformation to the system in the following ways (when the mechanical oscillators are non-deformed i.e. $q_{\textrm{m}1}=q_{\textrm{m}2}=1$.):
\begin{itemize}
	\item Either one of the cavities ($q_{\textrm{c}1}$ or $q_{\textrm{c}2}$) can be deformed independently, by keeping the other cavity non-deformed ($q_{\textrm{c}1}=1$ or $q_{\textrm{c}2}=1$),
	\item Both cavities can be deformed simultaneously by tuning the deformation strength either in the same ($q_{\textrm{c}1} = q_{\textrm{c}2} = q$) or distinct ($q_{\textrm{c}1} \ne q_{\textrm{c}2}$) manner. 
\end{itemize}
In this article we will be focussing particularly on the deformation of the first cavity keeping the second cavity non-deformed i.e. $q_{\textrm{c}1} \le 1$ and $q_{\textrm{c}2}=1$, and the mechanical oscillators are non-deformed; if not stated otherwise. 
\subsubsection{Deformed OMIT}
By fixing $q_{\textrm{c}2}=q_{\textrm{m}1}=q_{\textrm{m}2}=1$, and varying $q_{\textrm{c}1} = q$, in the range $0<q\le 1$ we have observed the appearance and behaviour of TWs. As a first observation it is noted that as we tend to take the system further from the non-deformed condition towards $q\rightarrow 0$, the symmetrical nature of the probe field behavior on the both side of $\beta=0$ gradually deminishes. The amplitude of the absorption peak exceeds maximum absorption limit (which is $2$) twice. For first time at $q=0.7$ the absorption coefficient reaches to $\approx2.43$, $2.16$ at $\beta/\kappa_1\approx-0.406$ and $-1.35$ respectively. For the second time at $q=0.3$ where the absorption peak takes the value $\approx8.5$ and $3.97$ at $\beta/\kappa_1\approx-0.3468$ and $-1.21$ respectively. On the other hand the amplitude of absorption for $\beta/\kappa_1>0$ decreases. Another observation in when we vary the deformation strength in range  {$0.5\le q<1$} is that, the width of the TWs which appear in between $\beta=\pm2$ tends to become wider towards $q\rightarrow0.5$.

Interestingly as we deform the system further i.e. $q<0.5$ a negative absorption peak appears in the region $-2<\beta/\kappa_1<0$. The absorption peak takes a value of $\approx-1.00$ for $q=0.3$.  {The appearance of negative absorption peaks in optomechanical systems is not new, in Ref.\cite{xin2018nonmarkov} the authors detected the appearance of negative absorption that might be due the non-Markovian nature of the environment. In our observation we detect the gain in probe field which is caused due to the photon backflow from the environment, leading to more transmitted energy than provided at $q=0.3$. The gain in probe field might indicate that, although we initially started with Markovian assumption at $q=1$; but as we vary the deformation parameter in range $0<q<0.5$, the photon backflow becomes more prominent. Which is responsible for the negative absorption peak.}

\begin{figure}[b!]
	\centering
	\includegraphics[width=0.7\linewidth]{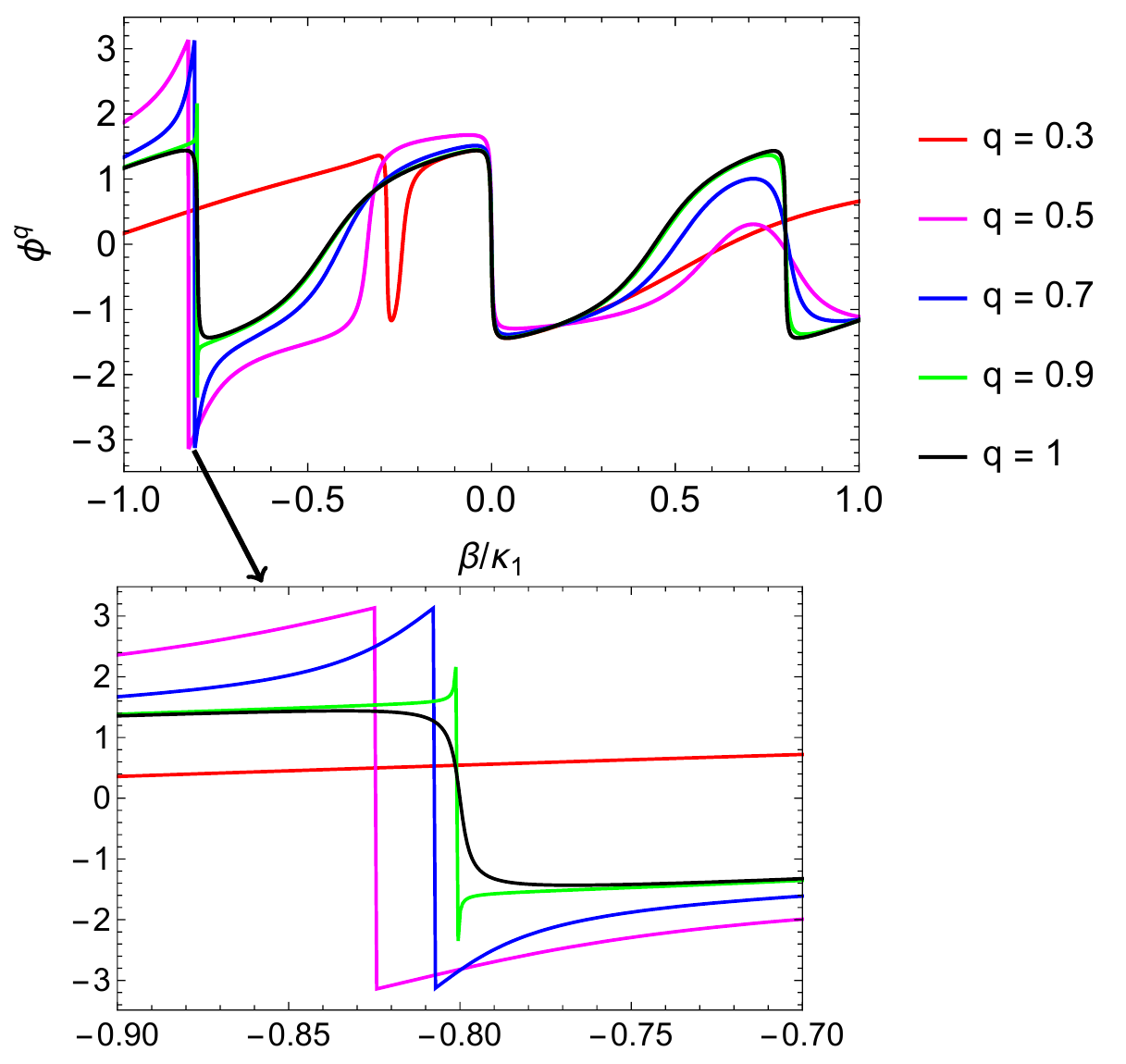}
	\caption{The deformed phase of the probe field as a function of $\beta/\kappa_1$.}
	\label{fig.phase_deformed_G1eqG2}
\end{figure}

\begin{figure}[b!]
	\centering
	\begin{subfigure}{0.5\textwidth}
		\centering
		\includegraphics[width=\linewidth]{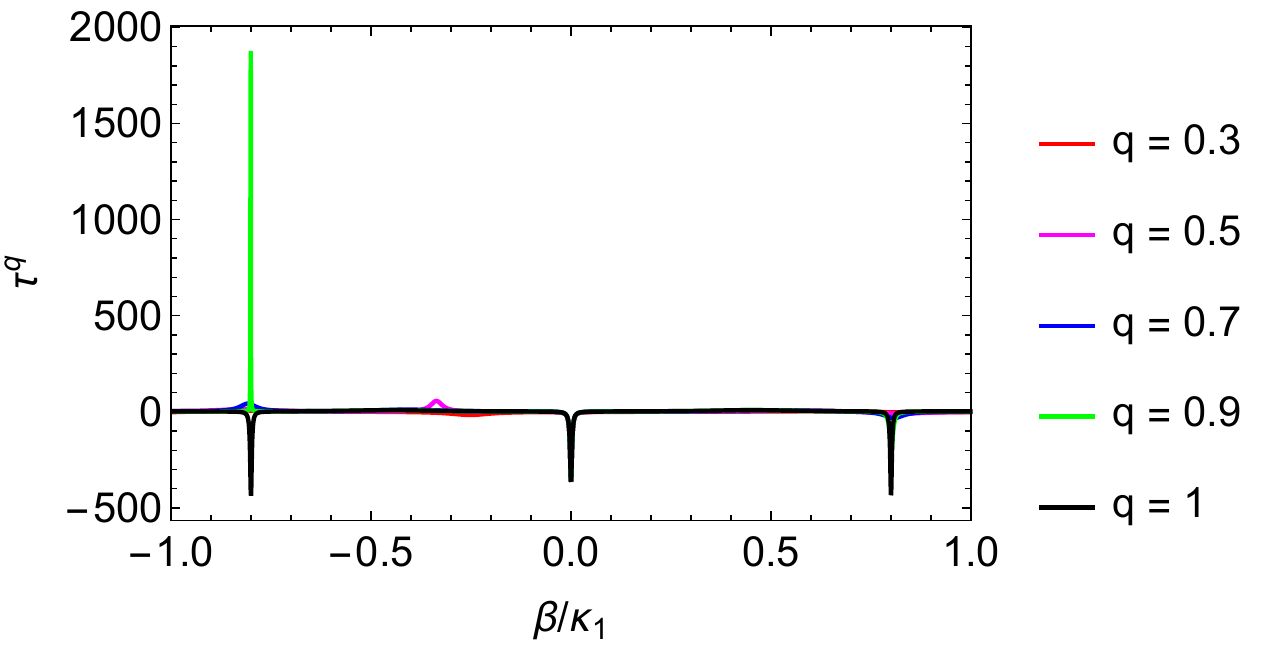}
		\caption{}
		\label{fig:delay_deformed_G1eqG2}
	\end{subfigure}%
	\begin{subfigure}{.54\textwidth}
		\centering
		\includegraphics[width=\linewidth]{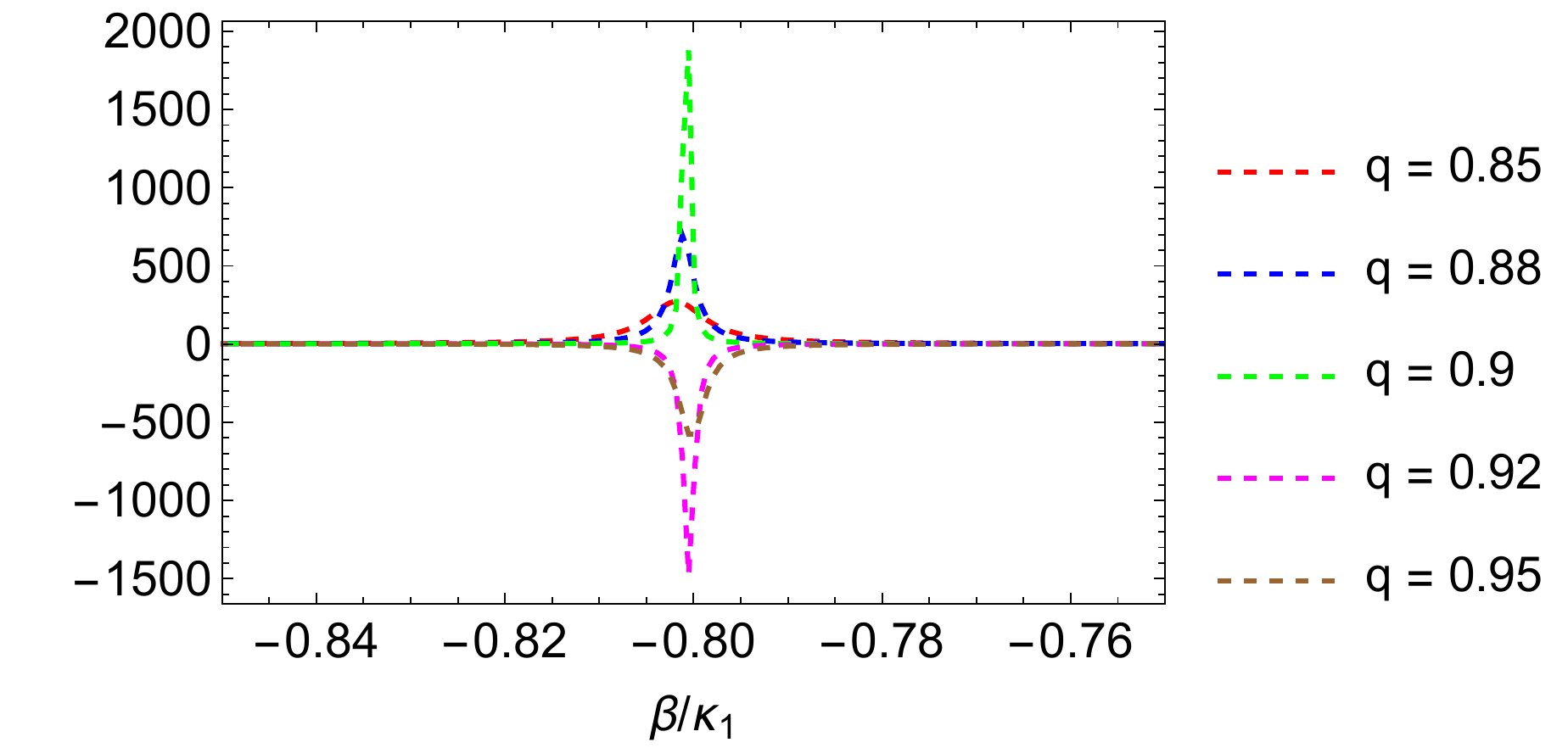}
		\caption{}
		\label{fig.condensed_delay_deformed_G1eqG2}
	\end{subfigure}
	\caption{In (a) we illustrate the deformed group delay of the probe field as a function of $\beta/\kappa_1$ in range $0.3\le q \le1$. In (b) we observe the tunability of positive and negative group delay around $q=0.9$. }
\end{figure}

\subsubsection{Deformed fast and slow light}
As a consequence of the deformation introduced in the system, we expect to observe a deformed phase and fast-slow light phenomena which is derived as follows
\begin{equation}
	\tau^q = \dfrac{\partial\phi^q(\omega_p)}{\partial\omega_p} = \dfrac{\partial\phi^q(\omega_p)}{\partial\beta},
	\label{eq:q-group-delay}
\end{equation}
where $\phi^q = \textrm{arg}\left[A_T^q\right]$ is the deformed phase of the
transmitted probe beam.

In Fig.(\ref{fig.phase_deformed_G1eqG2}) we illustrate the deformed phase diagram with normalized detuning $\beta/\kappa_1$, for $q=0.9$. As it has already discussed while observing the non-deformed phase a sharp change in phase indicates the possibility of manipulation of super and subluminal light. In Fig.(\ref{fig.phase_deformed_G1eqG2}) this sharp phase change is observed at  {$\beta\approx-0.8\kappa_1$}, which indicates that at that particular detuning value we can manipulate super and subluminal light which is depicted in Fig.(\ref{fig:delay_deformed_G1eqG2}). Compared to the highest possible subluminal peak for non-deformed case at $g=5\times10^3\kappa_1$, the subluminal peak for $q=0.9$ at $g=\kappa_1$ is approximately 5 times stronger.  {The appearance of the strong subluminal light at $q=0.9$ can be explained as follows. In Fig.(\ref{fig:delay_deformed_G1eqG2}) we can see that as we move from $q=1$ (the black line) to $q=0.9$ (the green line), due to induced nonlinearity the frequency of probe field increases (as predicted and illustrated in Fig.(\ref{fig.q-freq1})) and the absorption decreases, indicating prominence of normal dispersion phenomena i.e. the sharp enhancement of phase at $\beta/\kappa_1\approx-0.8$ in Fig.(\ref{fig:delay_deformed_G1eqG2}). Meanwhile in range $0<q<0.9$, the sharpness of the phase enhancement diminishes, indicating to the fact that normal dispersion becomes weaker.}
	
 {In Fig.(\ref{fig.condensed_delay_deformed_G1eqG2}), at $q=0.92$, we see an enhancement in superluminal light. Which indicates appearance of anomalous dispersion i.e. the phase of the probe field shifts from positive to negative value. Our observation also indicates that a slight change in deformation parameter may help us to tune from super to subluminal light and vice versa. As the TWs at $\beta/\kappa_1\approx-0.8$ vanishes as we shifts towards $q=0$, the tunability of fast and slow light becomes infeasible, which is illustrated in Fig.(\ref{fig.condensed_delay_deformed_G1eqG2}).}
\begin{figure}[t!]
	\centering
	\includegraphics[width=0.85\linewidth]{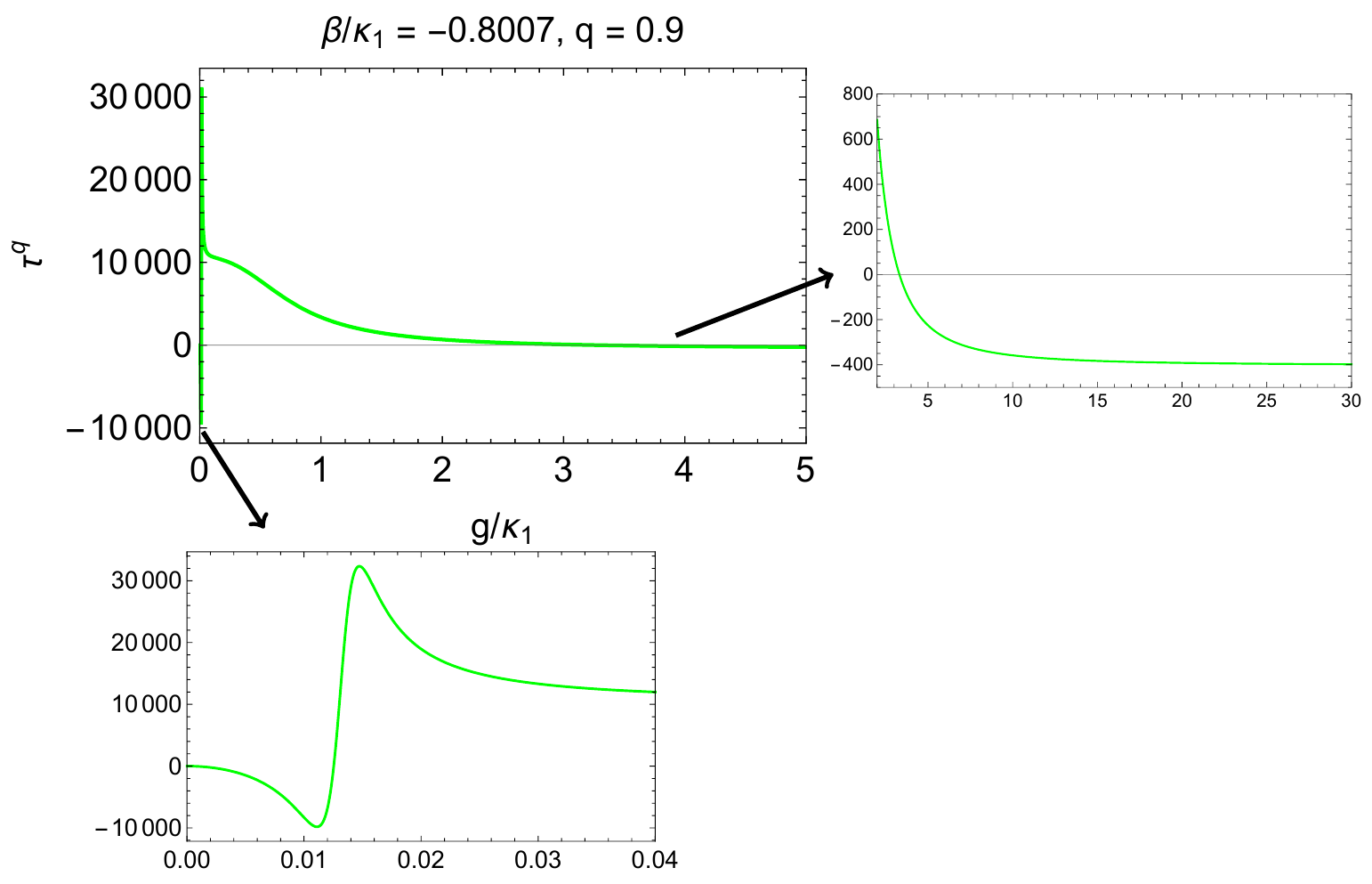}
	\caption{Deformed group delay with normalized tunnelling strength for $\beta = -0.8007\kappa_1$, and $q=0.9$.}
	\label{fig:delay_deformed_with_g}
\end{figure}

Using the cumulative knowledge of deformed and non-deformed cases we can expect an interplay of super and subluminal light is possible when we tune the tunnelling paramerter $g$ around  {$\beta\approx-0.8\kappa_1$} for $g=0.9$ which is depicted in Fig.(\ref{fig:delay_deformed_with_g}). In this figure using previous knowledge we fix the $q$ value and the $\beta$ to study the nature of group delay with tunnelling strength. The first observation one can infer from Fig.(\ref{fig:delay_deformed_with_g}) is that as the tunnelling strength increases the appearance of super and subluminal light deminishes. At a first glance although the group delay seems to saturate at null value when $g>2\kappa_1$,  but through magnification a transition from sub to superluminal light can be observe around $g\approx3.2\kappa_1$. The group delay saturates and the probe field shows an advancement of $400\mu\text{s}$ in the strong inter-cavity transmission strength region.

Meanwhile in the weak inter-cavity transmission strength regime particularly in the range $0.01<g/\kappa_1\approx0.015$, a very sharpe transition from super to subluminal light is observed. The probe field at $g\approx0.011\kappa_1$, shows an advancement in the order of $9.7m\text{s}$ and at $g\approx0.0146\kappa_1$ a delay of probe field is observed in the order of $32.3m\text{s}$. The observed characteristics of the group delay can be explained in the following way. At $q=0.9$ when $g$ is weak, then the photon in the second cavity impose a very weak perterbation to the photon in the first cavity, so the non-linearity in first cavity, that is induced in the from of deformation sustains as long as $g<3$. But when $g>3$, the influence of the photon in the second cavity destroys the induced nonlinearity of the first cavity. Hence the group delay tends to mimic the characteristics of an non-deformed cavity and the group delay saturates at a probe field advancement in the order of $400\mu\text{s}$.

\begin{figure}[t!]
	\centering
	\includegraphics[width=0.85\linewidth]{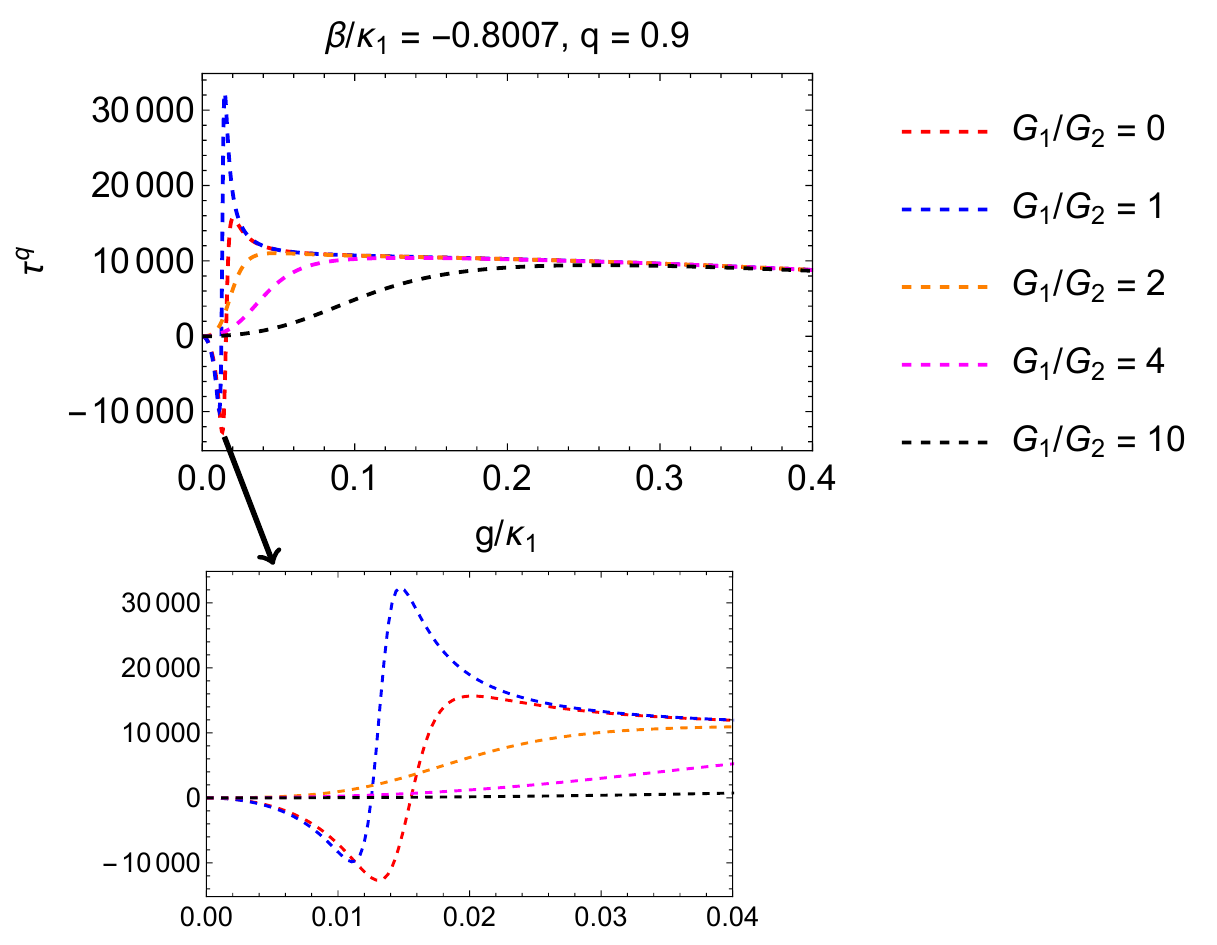}
	\caption{Deformed group delay with optomechanical coupling and tunnelling strength for $\beta^q = -0.809559\kappa_1^q$, and $q=0.9$.}
	\label{fig:delay_deformed_with_G}
\end{figure}

Finally, we investigate the group delay under different values of optomechanical coupling when $G_1/G_2\ge0$ as depicted in Fig.(\ref{fig:delay_deformed_with_G}).
Even when the first cavity is not interacting with the corresponding mechanical oscillator, we get a strong super and subluminal light in the order of milliseconds for $0<g/\kappa_1<0.02$. As $G_1$ approaches $G_2$, the strength of subluminal light enhances but superluminal light weakens gradually. Meanwhile when $G_1/G_2>1$, the sharpness of transition from super to subluminal light as well as the absolute strength of group delay deminishes.

\subsubsection{Fast and slow light above milliseconds}
By fixing the deformation strength at $q=0.9$ we show in Fig.(\ref{fig:strongest_subluminal}), Fig.(\ref{fig:strongest_superluminal}) that the probe field can be delayed and advanced above milliseconds respectively when fine tuning of normalized inter-cavity tunnelling strength and the detuning parameter is possible.

\section{Anyon models: $q$ as primitive root of unity}\label{sec:anyon}
More than four decades ago it has been pointed out \cite{leinaas1977theory} that in systems confined to two spatial dimensions, particles with general exchange statistics more than those of bosons and fermions, are possible. Particles with exchange statistics  {in between fermion and boson} were later named as \textit{anyon} by  {Wilczek} \cite{wilczek1982quantum}. In a general $q$ deformed system when the deformation strength is expressed as a primitive root of unity i.e. 
\begin{equation}
	q = \textrm{exp}\left(\dfrac{2\pi\textrm{i}}{l+2}\right),\label{eq:q-deformation-primitive-root}
\end{equation}
then this specific type of nonlinear model belongs to a general family of anyon models known as $SU(2)_l$ models. The mathematical framework that describes the $SU(2)_l$ models are called topological quantum field theory which provides tools to study physical processes in topological phases of matter.
\begin{figure}[tbh!]
	\centering
	\begin{subfigure}[b]{0.5\textwidth}
		\includegraphics[width=1\linewidth]{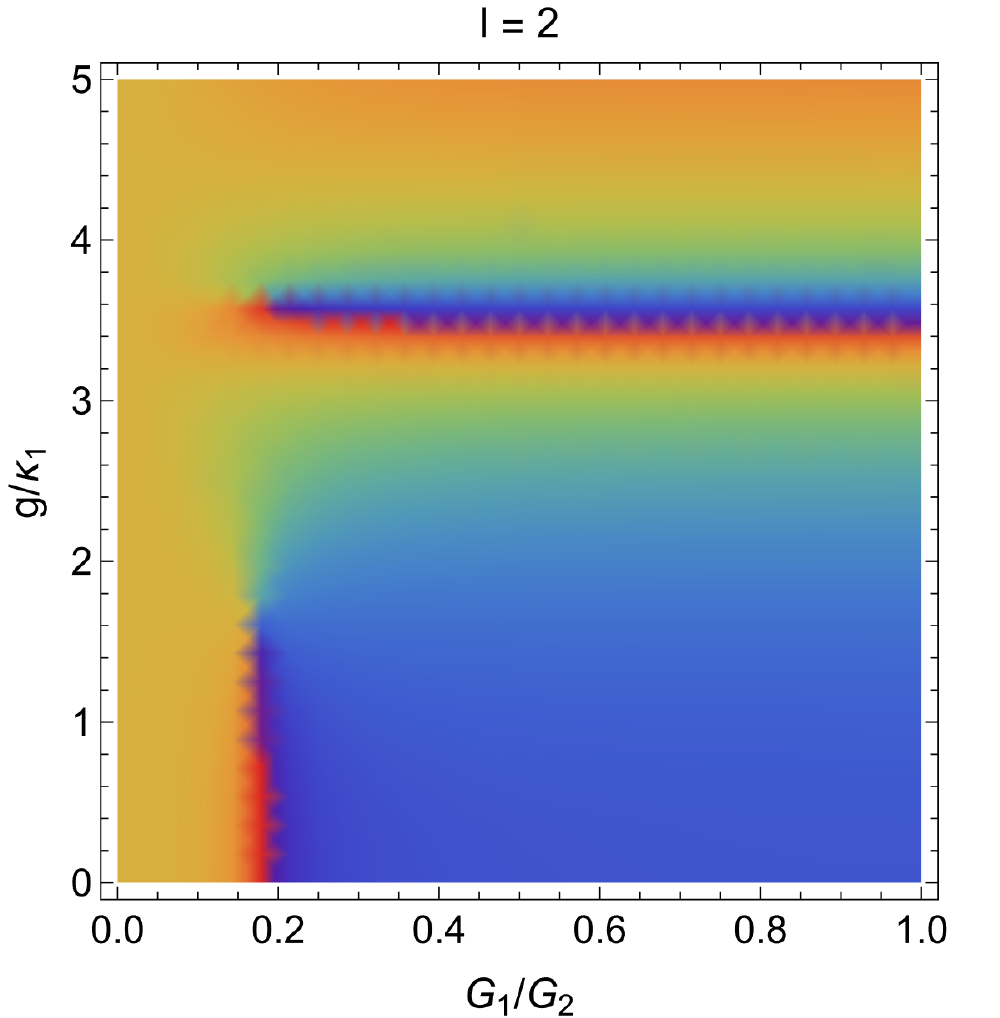}
		\caption{}
		\label{fig:ising_phase} 
	\end{subfigure}%
	\begin{subfigure}[b]{0.578\textwidth}
		\includegraphics[width=1\linewidth]{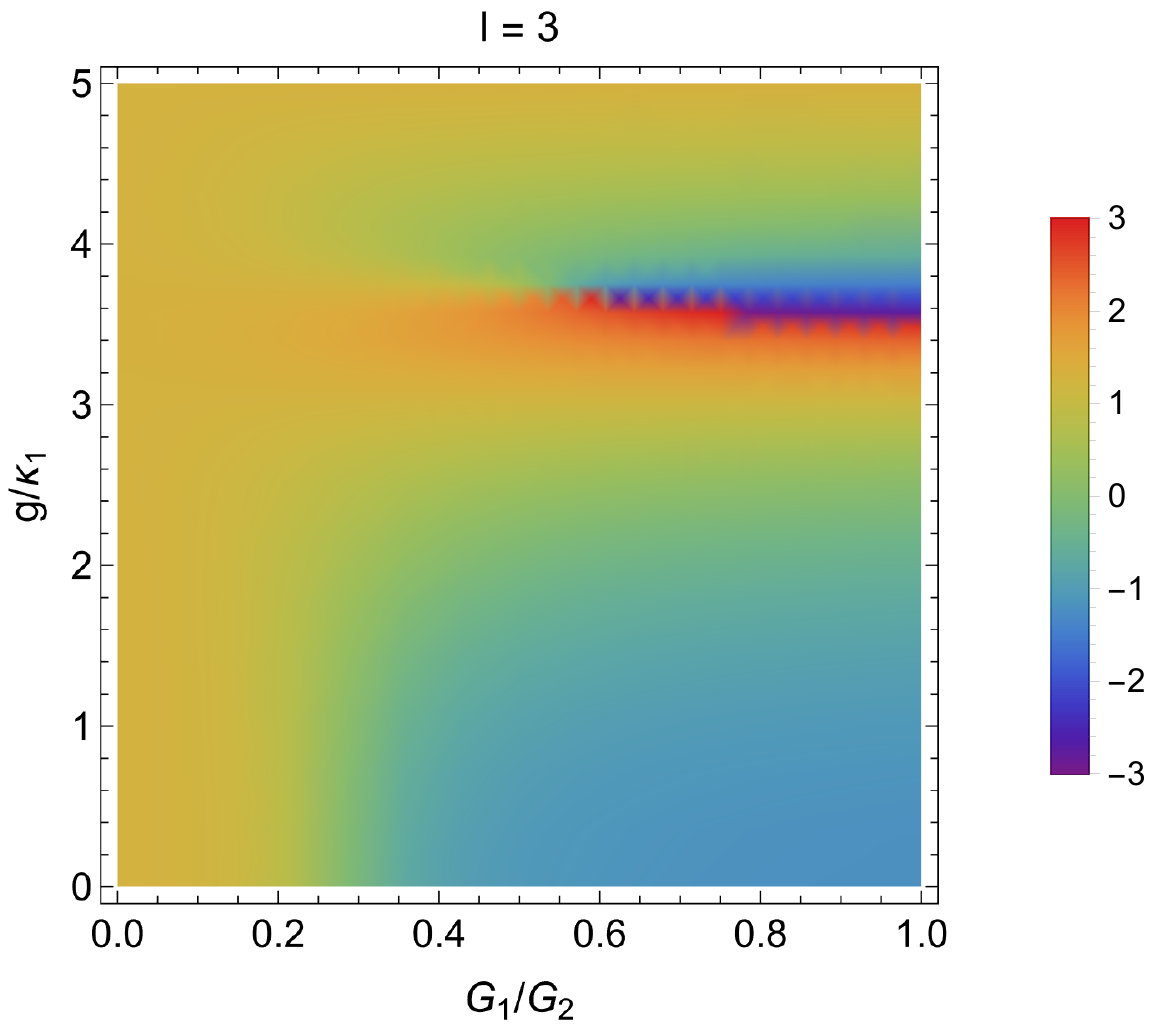}
		\caption{}
		\label{fig:fibonacci_phase}
	\end{subfigure}
	\caption{Illustration of variation of phase when the first cavity consists of (a) Ising anyons i.e. $l=2$, and for (b) Fibonacci anyons i.e. $l=3$. From the figure it is clearly visible that the appearance of rapid phase transition is more frequency in (a), which indicates better manipulation super and subluminal light.}
	\label{fig:anyon-phase}
\end{figure}

In this section we will be particularly focusing on \textit{Ising} and \textit{Fibonacci anyon} models by putting $l=2$ and $l=3$  {respectively in Eq.(\ref{eq:q-deformation-primitive-root})}. Although physical realization of Ising anyons is proposed in the form of Majorana fermion zero modes, but the evidence for Fibonacci anyons, has not been reported yet. The Fibonacci anyons are believed to exist as quasiparticles in Kondo systems \cite{komijani2020isolating} and in exotic quantum Hall fluids with filling fraction $12/5$ \cite{mong2017fibonacci,xia2004electron}.

By keeping the second cavity  {and} the mechanical oscillators non-deformed, we deform the first cavity setting $q_{\textrm{c}1}=q=\textrm{exp}\left({\dfrac{2\pi\textrm{i}}{l+2}}\right)$.  {This is indicating} that the $q$-deformed particles in the first cavity are behaving like Ising and Fibonacci anyons respectively. We observe the behavior of phase transition, finally fast and slow light of the probe field.

 {Fig.(\ref{fig:anyon-phase}) depicts the phase transition behavior of the probe field when the first cavity consists of Ising (see Fig.(\ref{fig:ising_phase})) and Fibonacci (see Fig.(\ref{fig:fibonacci_phase})) anyons. The appearance of rapid phase transition in probe field is more frequent and prominent when the first cavity consists of Ising anyons compared to Fibonacci anyons. The reason behind this observation can be described as follows. In case of the Ising anyons the increase in frequency of the probe field due to induced nonlinearity is stronger than the case when we Fibonacci anyons are considered (which can be deducted from Fig.(\ref{fig.q-freq1})). This observation directly indicates stronger normal dispersion, causing appearance of subluminal light. And the strength of subluminal peak is stronger in case of Ising anyons compared to Fibonacci ones as illustrated in the group delay plot in Fig.(\ref{fig:delay-ising}) and Fig.(\ref{fig:delay-fibonacci}) respectively.}

 { A more elaborated discussion is given as follows. For Ising anyons which is illustrated in figure Fig.(\ref{fig:ising_phase}), when the normalized inter-cavity coupling is $0 < g/k_1 < 0.8$, and when the $ G_1/G_2 < 0.2 $, the phase of the probe field increase sharply (transition from yellow to the red region), indicating strong normal dispersion behavior and appearance of strong subluminal peak. A slight increase in $g$, tends to steep decrease in the phase of the probe field towards negative value (transition red to deep blue region), which in turns indicates appearance of strong anomalous dispersion and we see strong superluminal light. This description also holds for Fibonacci anyons illustrated in Fig.(\ref{fig:fibonacci_phase}), but as the steepness of the variation of phase is low compared to Ising anyons so we observe weaker normal and anomalous dispersion, causing relatively weak super and subluminal light.}

\section{Summary and conclusions}\label{sec:conclusion}
As a summary, in this work, we present a comprehensive review on $q$-deformation and analytically study the tunability of optomechanical induced transparency in a $q$-deformed optomechanical system. The system contains two $q$-deformed cavities which are coupled to deformed mechanical oscillators, and with each other by tunneling parameter. We have particularly examined the case when the mechanical membranes and the second cavity is non-deformed but the first cavity is influenced by the deformation strength $q$. Under the influence of $q$-deformation the dynamics of the cavity depends upon $q$-number representation of bosonic number operator. When  {the deformation strength} is real and $q<1$, the deformation in the first cavity induces a nonlinearity to the optomechanical system. As a result, in the region $0<q<0.5$, the system experiences gain, which is caused by photon backflow from the environment to the first cavity. Moreover, we achieve a significant enhancement in the strength of fast and slow light phenomena in a scale of \textit{milliseconds} and even above by fixing the deformation of the first cavity at $q=0.9$ {, and by finely tuning the normalized inter-cavity coupling $g/\kappa_1$ and detuning $\beta/\kappa_1$}. In addition, to  {reduce} the gap between the mathematical $q$-deformation scheme and physically achievable systems, we define $q$ as primitive root of unity, indicating  {a class} anyon models. We investigate the phase transition and the fast-slow light behavior of the probe field under the two assumptions (1) the first cavity's $q$-deformed particles behave like Ising anyons at  {deformation strength $q=\textrm{exp}\left(\pi\textrm{i}/2\right)$} and (2)  {when the $q$-deformed particles mimics Fibonacci anyons at}  {$q=\text{exp}\left(2\pi\textrm{i}/5\right)$}. In both cases the probe field experiences enhanced rapidity in phase change,  {indicating towards more tunability of group delay and enhancement in fast-slow light compared to the non-deformed case}. Notably Ising anyons under the influence of same optomechanical parameters shows delay and advancement in probe field in the order of milliseconds and even in seconds. Our study introduces a new tool to not only investigate various cavity optomechanical systems, but to achieve enhancement over systems which do not necessitate $q$-deformation setup.

As an extension to this work, it would be interesting to see how the $q$-deformed system behaves when a two-level atomic system is introduced inside one or multiple cavities. Introducing atomic ensemble could make the width of the OMIT wider \cite{sohail2016optomechanically}, which may allow better manipulation of the deformed system. Meanwhile another obvious extension to this work could be to observe how the deformed system responds when we consider the strength of deformation in cavities are distinct (i.e. $q_{\textrm{c}1}\neq q_{\textrm{c}2}$). Moreover as our work only study Ising and Fibonacci anyon models when $l=2,3$ respectively, an investigation of the other classes of anyon models when $l>3$ \cite{ardonne2011microscopic,johansen2021fibonacci} is yet to be explored.

\section*{Acknowledgment}
The authors would like to acknowledge fruitful discussions with Marek Gluza. This work was supported by Polish National Center project 2019/33/B/ST6/02011.

\section*{Funding}
This work is supported by Polish National Center project 2019/33/B/ST6/02011.
\section*{Conflict of Interest}
The authors declare that they have no conflict of interest.
\section*{Data availability}
The datasets analysed during the current study are available from the corresponding author on reasonable request.

\bibliographystyle{unsrt}
\bibliography{reference_list}

\begin{thebibliography}{10}

\bibitem{sklyanin1982some}
Evgeny~Konstantinovich Sklyanin.
\newblock Some algebraic structures connected with the yang—baxter equation.
\newblock {\em Functional Analysis and its Applications}, 16(4):263--270, 1982.

\bibitem{kulish1983quantum}
Peter~P Kulish and N~Yu Reshetikhin.
\newblock Quantum linear problem for the sine-gordon equation and higher
  representations.
\newblock {\em Journal of Soviet Mathematics}, 23(4):2435--2441, 1983.

\bibitem{zamolodchikov1990factorized}
Alexander~B Zamolodchikov and Alexey~B Zamolodchikov.
\newblock Factorized s-matrices in two dimensions as the exact solutions of
  certain relativistic quantum field theory models.
\newblock In {\em Yang-Baxter Equation In Integrable Systems}, pages 82--120.
  World Scientific, 1990.

\bibitem{frohlich1992non}
Jurg Frohlich.
\newblock {\em Non-perturbative quantum field theory: Mathematical aspects and
  applications}, volume~15.
\newblock World Scientific, 1992.

\bibitem{moore1988polynomial}
Gregory Moore and Nathan Seiberg.
\newblock Polynomial equations for rational conformal field theories.
\newblock {\em Physics Letters B}, 212(4):451--460, 1988.

\bibitem{frenkel1988vertex}
Igor~B Frenkel and Naihuan Jing.
\newblock Vertex representations of quantum affine algebras.
\newblock {\em Proceedings of the National Academy of Sciences},
  85(24):9373--9377, 1988.

\bibitem{bernard1989vertex}
Denis Bernard.
\newblock Vertex operator representations of the quantum affine algebra
  $\mathcal{U}_q ({B}_r^{(1)})$.
\newblock {\em Letters in Mathematical Physics}, 17(3):239--245, 1989.

\bibitem{macfarlane1989q}
AJ~Macfarlane.
\newblock On q-analogues of the quantum harmonic oscillator and the quantum
  group ${SU}(2)q$.
\newblock {\em Journal of Physics A: Mathematical and General}, 22(21):4581,
  1989.

\bibitem{man1995deformed}
VI~Man'ko, G~Marmo, and F~Zaccaria.
\newblock Deformed field equations.
\newblock {\em Physics Letters A}, 197(2):95--99, 1995.

\bibitem{dey2015q}
Sanjib Dey.
\newblock Q-deformed noncommutative cat states and their nonclassical
  properties.
\newblock {\em Physical Review D}, 91(4):044024, 2015.

\bibitem{BAYINDIR2021105474}
Cihan Bayındır, Azmi~Ali Altintas, and Fatih Ozaydin.
\newblock Self-localized solitons of a q-deformed quantum system.
\newblock {\em Communications in Nonlinear Science and Numerical Simulation},
  92:105474, 2021.

\bibitem{altintas2014constructing}
Azmi~Ali Altintas, Fatih Ozaydin, Can Yesilyurt, Sinan Bugu, and Metin Arik.
\newblock Constructing quantum logic gates using q-deformed harmonic oscillator
  algebras.
\newblock {\em Quantum Information Processing}, 13(4):1035--1044, 2014.

\bibitem{altintas2020q}
Azmi~Ali Altintas, Fatih Ozaydin, and Cihan Bay{\i}nd{\i}r.
\newblock q-deformed three-level quantum logic.
\newblock {\em Quantum Information Processing}, 19(8):1--13, 2020.

\bibitem{PhysRevLett.98.030405}
D.~Vitali, S.~Gigan, A.~Ferreira, H.~R. B\"ohm, P.~Tombesi, A.~Guerreiro,
  V.~Vedral, A.~Zeilinger, and M.~Aspelmeyer.
\newblock Optomechanical entanglement between a movable mirror and a cavity
  field.
\newblock {\em Phys. Rev. Lett.}, 98:030405, Jan 2007.

\bibitem{arcizet2006radiationcooling}
Olivier Arcizet, P-F Cohadon, Tristan Briant, Michel Pinard, and Antoine
  Heidmann.
\newblock Radiation-pressure cooling and optomechanical instability of a
  micromirror.
\newblock {\em Nature}, 444(7115):71--74, 2006.

\bibitem{PhysRevLett.99.093901cooling}
I.~Wilson-Rae, N.~Nooshi, W.~Zwerger, and T.~J. Kippenberg.
\newblock Theory of ground state cooling of a mechanical oscillator using
  dynamical backaction.
\newblock {\em Phys. Rev. Lett.}, 99:093901, Aug 2007.

\bibitem{PhysRevLett.99.093902cooling}
Florian Marquardt, Joe~P. Chen, A.~A. Clerk, and S.~M. Girvin.
\newblock Quantum theory of cavity-assisted sideband cooling of mechanical
  motion.
\newblock {\em Phys. Rev. Lett.}, 99:093902, Aug 2007.

\bibitem{teufel2011sidebandcooling}
John~D Teufel, Tobias Donner, Dale Li, Jennifer~W Harlow, MS~Allman, Katarina
  Cicak, Adam~J Sirois, Jed~D Whittaker, Konrad~W Lehnert, and Raymond~W
  Simmonds.
\newblock Sideband cooling of micromechanical motion to the quantum ground
  state.
\newblock {\em Nature}, 475(7356):359--363, 2011.

\bibitem{PhysRevA.78.032316entanglement}
C.~Genes, A.~Mari, P.~Tombesi, and D.~Vitali.
\newblock Robust entanglement of a micromechanical resonator with output
  optical fields.
\newblock {\em Phys. Rev. A}, 78:032316, Sep 2008.

\bibitem{PhysRevA.89.014302entanglement}
Jie-Qiao Liao, Qin-Qin Wu, and Franco Nori.
\newblock Entangling two macroscopic mechanical mirrors in a two-cavity
  optomechanical system.
\newblock {\em Phys. Rev. A}, 89:014302, Jan 2014.

\bibitem{PhysRevLett.102.020501entanglement}
K.~Hammerer, M.~Aspelmeyer, E.~S. Polzik, and P.~Zoller.
\newblock Establishing einstein-poldosky-rosen channels between nanomechanics
  and atomic ensembles.
\newblock {\em Phys. Rev. Lett.}, 102:020501, Jan 2009.

\bibitem{PhysRevLett.88.120401entanglement}
Stefano Mancini, Vittorio Giovannetti, David Vitali, and Paolo Tombesi.
\newblock Entangling macroscopic oscillators exploiting radiation pressure.
\newblock {\em Phys. Rev. Lett.}, 88:120401, Mar 2002.

\bibitem{weis2010OMIT}
Stefan Weis, R{\'e}mi Rivi{\`e}re, Samuel Del{\'e}glise, Emanuel Gavartin,
  Olivier Arcizet, Albert Schliesser, and Tobias~J Kippenberg.
\newblock Optomechanically induced transparency.
\newblock {\em Science}, 330(6010):1520--1523, 2010.

\bibitem{karuza2013OMIT}
M~Karuza, C~Biancofiore, M~Bawaj, C~Molinelli, M~Galassi, R~Natali, P~Tombesi,
  G~Di~Giuseppe, and D~Vitali.
\newblock Optomechanically induced transparency in a membrane-in-the-middle
  setup at room temperature.
\newblock {\em Physical Review A}, 88(1):013804, 2013.

\bibitem{PhysRevA.90.043825OMIT}
Peng-Cheng Ma, Jian-Qi Zhang, Yin Xiao, Mang Feng, and Zhi-Ming Zhang.
\newblock Tunable double optomechanically induced transparency in an
  optomechanical system.
\newblock {\em Phys. Rev. A}, 90:043825, Oct 2014.

\bibitem{harris1990EIT}
Stephen~E Harris, JE~Field, and A~Imamo{\u{g}}lu.
\newblock Nonlinear optical processes using electromagnetically induced
  transparency.
\newblock {\em Physical Review Letters}, 64(10):1107, 1990.

\bibitem{boller1991EIT}
K-J Boller, A~Imamo{\u{g}}lu, and Stephen~E Harris.
\newblock Observation of electromagnetically induced transparency.
\newblock {\em Physical Review Letters}, 66(20):2593, 1991.

\bibitem{fleischhauer2005EIT}
Michael Fleischhauer, Atac Imamo{\u{g}}lu, and Jonathan~P Marangos.
\newblock Electromagnetically induced transparency: Optics in coherent media.
\newblock {\em Reviews of Modern Physics}, 77(2):633, 2005.

\bibitem{thevenaz2008slow}
Luc Th{\'e}venaz.
\newblock Slow and fast light in optical fibres.
\newblock {\em Nature Photonics}, 2(8):474--481, 2008.

\bibitem{baba2008slow}
Toshihiko Baba.
\newblock Slow light in photonic crystals.
\newblock {\em Nature Photonics}, 2(8):465--473, 2008.

\bibitem{liao2020slow}
Qinghong Liao, Xing Xiao, Wenjie Nie, and Nanrun Zhou.
\newblock Transparency and tunable slow-fast light in a hybrid cavity
  optomechanical system.
\newblock {\em Optics express}, 28(4):5288--5305, 2020.

\bibitem{hussain2020fsstudy}
Anwar Hussain, Muqaddar Abbas, et~al.
\newblock Double transparency with slow and fast light in an optomechanical
  system.
\newblock {\em Optics Communications}, 461:125284, 2020.

\bibitem{liu2019nonreciprocalstudy}
Jun-Hao Liu, Ya-Fei Yu, and Zhi-Ming Zhang.
\newblock Nonreciprocal transmission and fast-slow light effects in a cavity
  optomechanical system.
\newblock {\em Optics express}, 27(11):15382--15390, 2019.

\bibitem{wang2019mechanical}
Bao Wang, Zeng-Xing Liu, Cui Kong, Hao Xiong, and Ying Wu.
\newblock Mechanical exceptional-point-induced transparency and slow light.
\newblock {\em Optics express}, 27(6):8069--8080, 2019.

\bibitem{KUNDU2021168465}
Akash Kundu, Chao Jin, and Jia-Xin Peng.
\newblock Optical response of a dual membrane active–passive optomechanical
  cavity.
\newblock {\em Annals of Physics}, 429:168465, 2021.

\bibitem{sun1989q}
Chang-Pu Sun and Hong-Chen Fu.
\newblock The q-deformed boson realisation of the quantum group ${SU}(n)q$ and
  its representations.
\newblock {\em Journal of Physics A: Mathematical and General}, 22(21):L983,
  1989.

\bibitem{biedenharn1989quantum}
LC~Biedenharn.
\newblock The quantum group ${SU}_q (2)$ and a q-analogue of the boson
  operators.
\newblock {\em Journal of Physics A: Mathematical and General}, 22(18):L873,
  1989.

\bibitem{bonatsos1999quantum}
Dennis Bonatsos and C~Daskaloyannis.
\newblock Quantum groups and their applications in nuclear physics.
\newblock {\em Progress in Particle and Nuclear Physics}, 43:537--618, 1999.

\bibitem{floreanini1991q}
Roberto Floreanini and Luc Vinet.
\newblock q-orthogonal polynomials and the oscillator quantum group.
\newblock {\em Letters in Mathematical Physics}, 22(1):45--54, 1991.

\bibitem{van1992q}
J~Van~der Jeugt.
\newblock The q-boson operator algebra and q-hermite polynomials.
\newblock {\em Letters in Mathematical Physics}, 24(4):267--274, 1992.

\bibitem{chang1992q}
Zhe Chang, Han-Ying Guo, and Hong Yan.
\newblock The q-{Hermite} polynomial and the representations of {Heisenberg}
  and quantum {Heisenberg} algebras.
\newblock {\em Journal of Physics A: Mathematical and General}, 25(6):1517,
  1992.

\bibitem{exton1983q}
Harold Exton.
\newblock {\em q-Hypergeometric functions and applications}.
\newblock Halsted Press, 1983.

\bibitem{gray1990completeness}
Robert~W Gray and Charles~A Nelson.
\newblock A completeness relation for the q-analogue coherent states by
  q-integration.
\newblock {\em Journal of Physics A: Mathematical and General}, 23(18):L945,
  1990.

\bibitem{bracken1991q}
AJ~Bracken, DS~McAnally, RB~Zhang, and MD~Gould.
\newblock A q-analogue of {Bargmann} space and its scalar product.
\newblock {\em Journal of Physics A: Mathematical and General}, 24(7):1379,
  1991.

\bibitem{floratos1990quantum}
Emmanuel~G Floratos and TN~Tomaras.
\newblock A quantum mechanical analogue for the q-oscillator.
\newblock {\em Physics Letters B}, 251(1):163--166, 1990.

\bibitem{chaichian1990quantum}
Masud Chaichian, D~Ellinas, and P~Kulish.
\newblock Quantum algebra as the dynamical symmetry of the deformed
  jaynes-cummings model.
\newblock {\em Physical Review Letters}, 65(8):980, 1990.

\bibitem{man1993physical}
VI~Man'ko, G~Marmo, S~Solimeno, and F~Zaccaria.
\newblock Physical nonlinear aspects of classical and quantum q-oscillators.
\newblock {\em International Journal of Modern Physics A}, 8(20):3577--3597,
  1993.

\bibitem{lin2010coherent}
Qiang Lin, Jessie Rosenberg, Darrick Chang, Ryan Camacho, Matt Eichenfield,
  Kerry~J Vahala, and Oskar Painter.
\newblock Coherent mixing of mechanical excitations in nano-optomechanical
  structures.
\newblock {\em Nature Photonics}, 4(4):236--242, 2010.

\bibitem{zheng2012controllable}
Can Zheng, Xiaoshun Jiang, Shiyue Hua, Long Chang, Guanyu Li, Huibo Fan, and
  Min Xiao.
\newblock Controllable optical analog to electromagnetically induced
  transparency in coupled high-q microtoroid cavities.
\newblock {\em Optics express}, 20(16):18319--18325, 2012.

\bibitem{sohail2016optomechanically}
Amjad Sohail, Yang Zhang, Jun Zhang, and Chang-shui Yu.
\newblock Optomechanically induced transparency in multi-cavity optomechanical
  system with and without one two-level atom.
\newblock {\em Scientific Reports}, 6(1):1--8, 2016.

\bibitem{aspelmeyer2014cavity}
Markus Aspelmeyer, Tobias~J Kippenberg, and Florian Marquardt.
\newblock Cavity optomechanics.
\newblock {\em Reviews of Modern Physics}, 86(4):1391, 2014.

\bibitem{xin2018nonmarkov}
Chun~Yu Xin, Shu~Sheng Meng, and YH~Zhou.
\newblock Non-markovian effect in optomechanical system.
\newblock {\em International Journal of Theoretical Physics}, 57(6):1659--1670,
  2018.

\bibitem{leinaas1977theory}
Jon~M Leinaas and Jan Myrheim.
\newblock On the theory of identical particles.
\newblock {\em Il Nuovo Cimento B (1971-1996)}, 37(1):1--23, 1977.

\bibitem{wilczek1982quantum}
Frank Wilczek.
\newblock Quantum mechanics of fractional-spin particles.
\newblock {\em Physical review letters}, 49(14):957, 1982.

\bibitem{komijani2020isolating}
Yashar Komijani.
\newblock Isolating kondo anyons for topological quantum computation.
\newblock {\em Physical Review B}, 101(23):235131, 2020.

\bibitem{mong2017fibonacci}
Roger~SK Mong, Michael~P Zaletel, Frank Pollmann, and Zlatko Papi{\'c}.
\newblock Fibonacci anyons and charge density order in the 12/5 and 13/5
  quantum hall plateaus.
\newblock {\em Physical Review B}, 95(11):115136, 2017.

\bibitem{xia2004electron}
JS~Xia, W~Pan, CLet Vicente, ED~Adams, NS~Sullivan, HL~Stormer, DC~Tsui,
  LN~Pfeiffer, KW~Baldwin, and KW~West.
\newblock Electron correlation in the second landau level: A competition
  between many nearly degenerate quantum phases.
\newblock {\em Physical review letters}, 93(17):176809, 2004.

\bibitem{ardonne2011microscopic}
Eddy Ardonne, Jan Gukelberger, Andreas~WW Ludwig, Simon Trebst, and Matthias
  Troyer.
\newblock Microscopic models of interacting yang--lee anyons.
\newblock {\em New Journal of Physics}, 13(4):045006, 2011.

\bibitem{johansen2021fibonacci}
Emil~G{\'e}netay Johansen and Tapio Simula.
\newblock Fibonacci anyons versus majorana fermions: A monte carlo approach to
  the compilation of braid circuits in su (2) k anyon models.
\newblock {\em PRX Quantum}, 2(1):010334, 2021.

\end{thebibliography}

\appendix

\section{Derivation of deformed frequency}\label{sec:q-freq}
The commutator relations used for the deformed system are as follows:
\begin{itemize}
	\item $\left[\hat{O}^q, \hat{O}^{q\dagger}\right] = [\hat{N}_O+1] - [\hat{N}_O] = \dfrac{q^{N_O+1}+q^{-N_O}}{1+q}$,
	\item $\left[\hat{Q}_{O^{q}}, \hat{P}_{O^{q}}\right] = i\left[\hat{O}^q, \hat{O}^{q\dagger}\right] = i\dfrac{q^{N_O+1}+q^{-N_O}}{1+q}$.
	\item $[\hat{N}_O+1] + [\hat{N}_O] = \dfrac{q^{N_O+1}-q^{-N_O}}{q-1}$ 
\end{itemize}
From equation (\ref{eq:non-deformed-hamiltonian-JC-form}) the dynamics of the bosonic operators given by,
\begin{equation}
	 {\frac{d\hat{O}}{dt}} =-i\omega\hat{O}\;\;\;;\;\;\; {\frac{d\hat{O}^\dagger}{dt}} =-i\omega\hat{O}^\dagger\label{eq:dynamics-QHO-hamiltonian}
\end{equation}
following the same way using Heisenberg equation, the dynamics of $q$-deformed bosonic operator defined by Hamiltonian (\ref{eq:$q$-deformed-hamiltonian}),
%\begin{widetext}
\begin{align}
	\frac{d\hat{O}}{dt}&=\hat{H}_{q\rightarrow\text{real}}\hat{O}-\hat{O}\hat{H}_{q\rightarrow\text{real}}=i\omega\dfrac{\text{sinh}\left[k\left(\hat{O}^\dagger\hat{O}\right)\right]}{\text{sinh}k}\hat{O}-i\omega\hat{O}\dfrac{\text{sinh}\left[k\left(\hat{O}^\dagger\hat{O}\right)\right]}{\text{sinh}k},\nonumber\\
	&=\dfrac{i\omega}{\text{sinh}k}\left[\left(k\hat{O}^\dagger\hat{O}+\dfrac{k^3}{3!}\left(\hat{O}^\dagger\hat{O}\right)^3+\ldots\right)\hat{O}-\hat{O}\left(k\hat{O}^\dagger\hat{O}+\dfrac{k^3}{3!}\left(\hat{O}^\dagger\hat{O}\right)^3+\ldots\right)\right],\nonumber\\
	&=\dfrac{i\omega k}{\text{sinh}k}\left[-\hat{O}-\dfrac{k^2}{3!}3\left(\hat{O}^\dagger\hat{O}\right)^2\hat{O}+\ldots\right]=-i\omega\dfrac{k}{\text{sinh}k}\left[1+\dfrac{k^2}{2!}\left(\hat{O}^\dagger\hat{O}\right)^2+\ldots\right],\nonumber
\end{align}
%\end{widetext}
form where we finally arrive at,
\begin{equation}
	\hat{O}=-i\omega_q\hat{O}\label{eq:q-dynamics-QHO-hamiltonian}\;\;\;;\;\;\;\hat{O}^\dagger=-i\omega_q\hat{O},
\end{equation}
where,
\begin{equation}
	\omega_q=\omega\dfrac{k\text{cosh}\left[k\left(\hat{O}^\dagger\hat{O}\right)\right]}{\text{sinh}k},\label{eq:appendixa-q-angular-momentum}
\end{equation}
which represents the orbital dependence frequency after passing through $q$-deformation.

\section{Detailed derivation of dynamics}\label{sec:method}
The steady state solutions by Equations (\ref{eq-ave1}) to (\ref{eq-ave2}) are given by,
\begin{eqnarray}
	&&\bar{c}_1^{q_{\textrm{c}1}} = \varepsilon_c / \left[\left(\kappa_1^{q_{\textrm{c}1}} + \dfrac{g^{2}}{\kappa_2^{q_{\textrm{c}2 }} + i\tilde{\Delta}_2}\right)+i\tilde{\Delta}_1\right]\label{eq:steady-c1},\\
	&&\bar{c}_2^{q_{\textrm{c}2}} = -\dfrac{ig\bar{c}_1^{q_{\textrm{c}1}}}{\kappa_2^{q_{\textrm{c}2}}+i\tilde{\Delta}_2},\\
	&& \tilde{\Delta}_i = \Delta_i - g_{\textrm{om}i}\bar{\alpha}_i,\\
	&&\bar{\alpha}_i = \dfrac{2\omega_{\textrm{m}i}g_{\textrm{om}i}}{\left(\gamma_{\textrm{m}i}^{{q_{\textrm{m}i}}}\right)^2 + \omega_{\textrm{m}i}^{^2}}\lvert \bar{c}_i^{q_{\textrm{c}i}} \rvert^2\label{eq:appendix:steady-alpha}.
\end{eqnarray}
 If we substitute $\langle \hat{O}(t) \rangle = \bar{O} + \delta\hat{O}(t)$ ($\hat{O} = \hat{c}_i^{q_{\textrm{c}i}},\;\hat{b}_i^{q_{\textrm{m}i}}, \;\;\;i=1,2$) in equations (\ref{eq-ave1}) to (\ref{eq-ave2}) one get an array of time dependent equation corresponding to $\delta\hat{O}(t)$ which should include the probing field term $\varepsilon_pe^{-i\Delta t}$. These equations can be given as follows
\begin{eqnarray}
	&&  {\frac{d}{dt}\delta c_1^{q_{\textrm{c}1}}} = \left[-\left(\kappa_1^{q_{\textrm{c}1}} + i\tilde{\Delta}_1\right)\delta\hat{c}_1^{q_{\textrm{c}1}}+ iG_1^{q_{\textrm{c}1}}\left(\delta\hat{b}_1^{q_{\textrm{m}1}} + \left(\delta\hat{b}_1^{q_{\textrm{m}1}}\right)^* \right) - ig\delta\hat{c}_2^{q_{\textrm{c}2}}+\varepsilon_pe^{-i\Delta t}\right]\chi_{N_{\textrm{c}1}}^{q_{\textrm{c}1}}\label{eq: time-dynamics1}\nonumber,\\
	&&  {\frac{d}{dt}\delta c_2^{q_{\textrm{c}2}}} = \left[-\left(\kappa_2^{q_{\textrm{c}2}} + i\tilde{\Delta}_2\right)\delta\hat{c}_2^{q_{\textrm{c}2}}+ iG_2^{q_{\textrm{c}2}}\left(\delta\hat{b}_2^{q_{\textrm{m}2}} + \left(\delta\hat{b}_2^{q_{\textrm{m}2}}\right)^* \right) - ig\delta\hat{c}_1^{q_{\textrm{c}1}}\right]\chi_{N_{c2}}^{q_{\textrm{c}2}}\nonumber,\\
	&&  {\frac{d}{dt}\delta b_i^{q_{\textrm{m}i}}} = \left[-\left(\gamma_{mi}^{q_{\textrm{m}i}}+ i\omega_{mi}\right)\delta\hat{b}_i^{q_{\textrm{m}i}}+ i\left(G_i^{q_{\textrm{c}i}}\left(\delta\hat{c}_i^{q_{\textrm{c}i}}\right)^* + \left(G_i^{q_{\textrm{c}i}}\right)^*\delta\hat{c}_i^{q_{\textrm{c}i}}\right)\right]\chi_{N_{\textrm{m}i}}^{q_{\textrm{m}i}}\label{eq: time-dynamics2}\nonumber,
\end{eqnarray}
\\
for $i =1,2$, $G_i^{q_{\textrm{c}i}} = g_{\textrm{om}i}\bar{c}_i^{q_{\textrm{c}i}}$  {and ``$*$" denoted complex conjugate}. To gain more physical we observe the dynamics of the system in resolved sideband regime. In this regime $\omega_{mi}\gg\kappa_i$ and $\tilde{\Delta}_i = \omega_{\text{m}i}$. In such regime, the lower sideband far off resonance can be safely neglected which means, $\delta\hat{O}(t) \approx \hat{O}_{-}e^{-i\Delta t}$ i.e. $\hat{O}_+ = 0$. Thus the equation from (\ref{eq: time-dynamics1}) to (\ref{eq: time-dynamics2}) can be rewritten as
\begin{eqnarray}
	&& 0 = -\left(\kappa_1^{q_{\textrm{c}1}}-i\beta^{q_{\textrm{c}1}}_{N_{c1}}\right)\hat{c}_{1,-}^{q_{\textrm{c}1}}+ iG_1^{q_{\textrm{c}1}}\hat{b}_{1,-}^{q_{\textrm{m}1}}- ig\hat{c}_{2,-}^{q_{\textrm{c}2}} + \varepsilon_p,\label{eq:transmission-gen-1}\\
	&& 0 = -\left(\kappa_2^{q_{\textrm{c}2}}-i\beta^{q_{\textrm{c}2}}_{N_{c2}}\right)\hat{c}_{2,-}^{q_{\textrm{c}2}} + iG_2^{q_{\textrm{c}2}}\hat{b}_{2,-}^{q_{\textrm{m}2}} - ig\hat{c}_{1,-}^{q_{\textrm{c}1}},\\
	&& 0 = -\left(\gamma_{\textrm{m}i}^{q_{\textrm{m}i}} - i\beta^{q_{\textrm{m}i}}_{N_{\text{m}i}}\right)\hat{b}_{i,-}^{q_{\textrm{m}i}} + i\left(G_i^{q_{\textrm{c}i}}\right)^*\hat{c}_{i,-}^{q_{\textrm{c}i}}\label{eq:transmission-gen-2},
\end{eqnarray}
where $\beta^{q_{\textrm{c}i}}_{N_{\textrm{c}i}} = \Delta^{q_{\textrm{c}i}}_{N_{\textrm{c}i}} - \omega_{\text{m}i}$, and $\beta_{N_{\text{m}i}}^{q_{\textrm{m}i}} = \Delta_{N_{\text{m}i}}^{q_{\textrm{m}i}} - \omega_{\text{m}i} $; where  $\Delta_{N_{\textrm{c}i}}^{q_{\textrm{c}i}} = \Delta/\chi_{N_{\textrm{c}i}}^{q_{\textrm{c}i}}$, and $\Delta_{N_{\textrm{m}i}}^{q_{\textrm{m}i}} = \Delta/\chi_{N_{\textrm{m}i}}^{q_{\textrm{m}i}}$.

\section{Illustrations}
\begin{figure}[tbh!]
	\centering
	\begin{subfigure}[b]{0.7\textwidth}
		\includegraphics[width=1\linewidth]{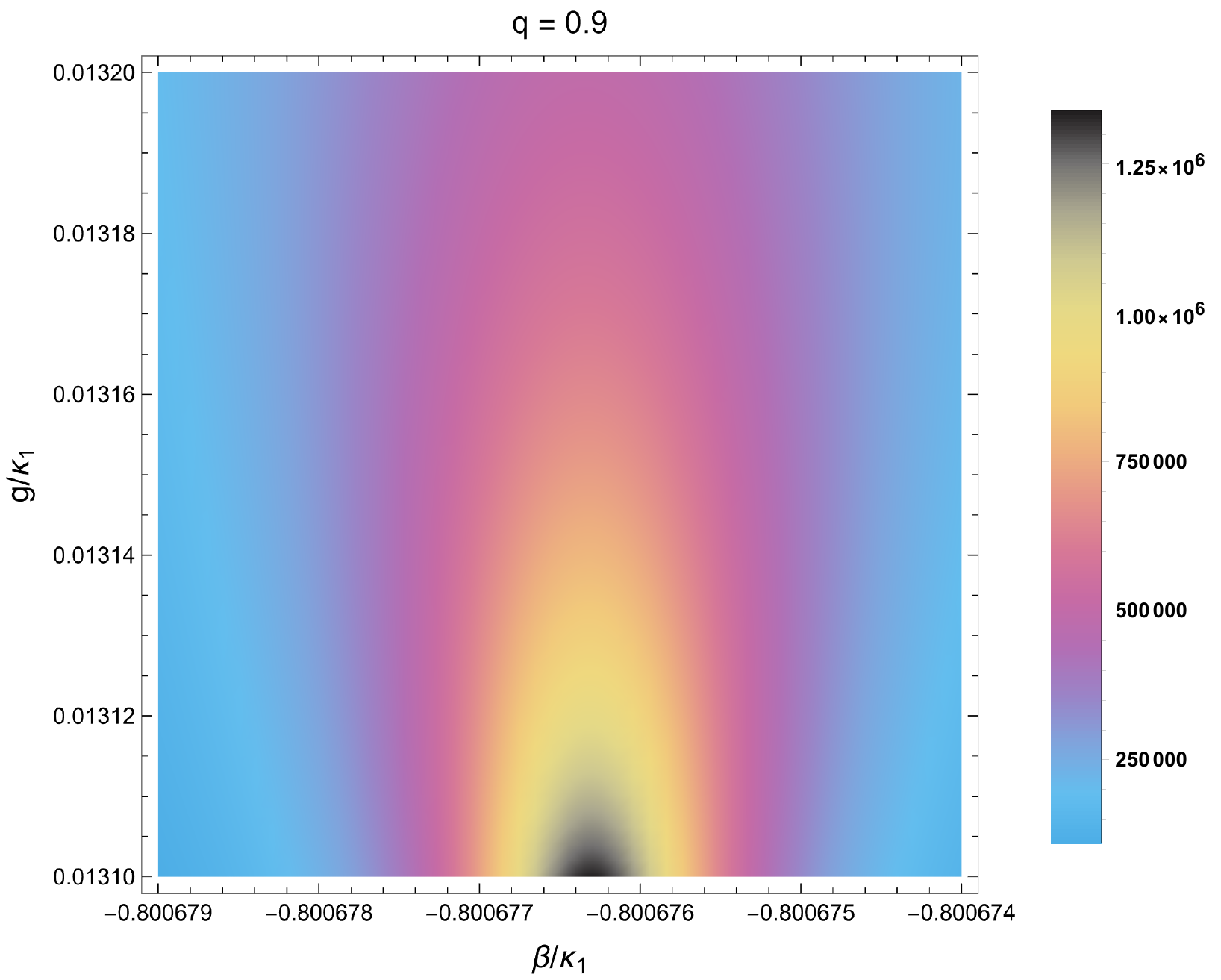}
		\caption{}
		\label{fig:strongest_subluminal}
	\end{subfigure}
	\begin{subfigure}[b]{0.7\textwidth}
		\includegraphics[width=1\linewidth]{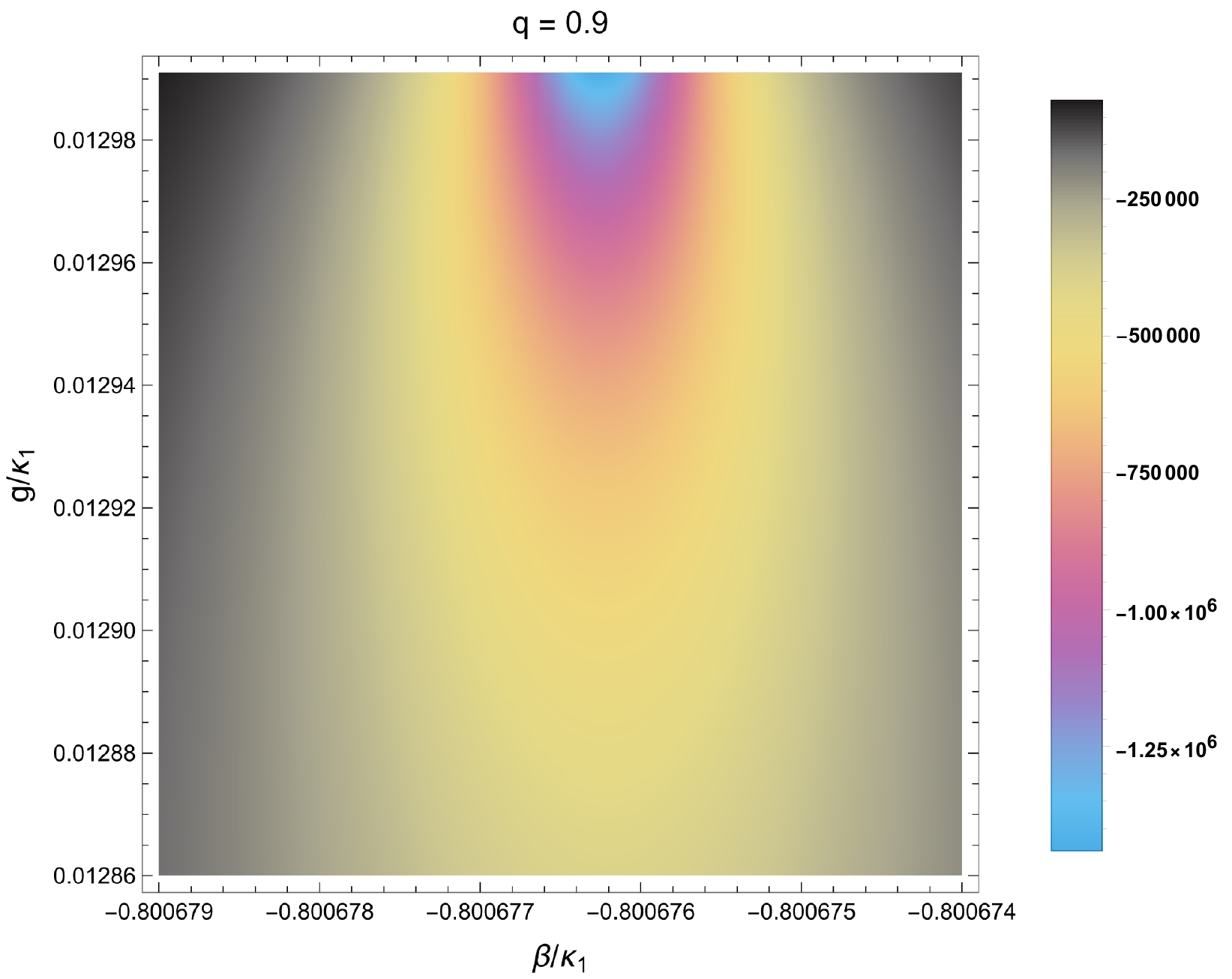}
		\caption{}
		\label{fig:strongest_superluminal}
	\end{subfigure}
	\caption{Illustration of variation of group delay in the probe field with normalized detuning (x-axis) and normalized inter-cavity transmission strength. In (a) we observe subluminal and in (b) we observe superluminal light above milliseconds for $q$ = $0.9$, and $\omega_{\text{m}1}=\omega_{\text{m}2}$.}
	\label{fig:deformed-delay}
\end{figure}

\begin{figure}[t!]
	\centering
	\includegraphics[width=0.84\linewidth]{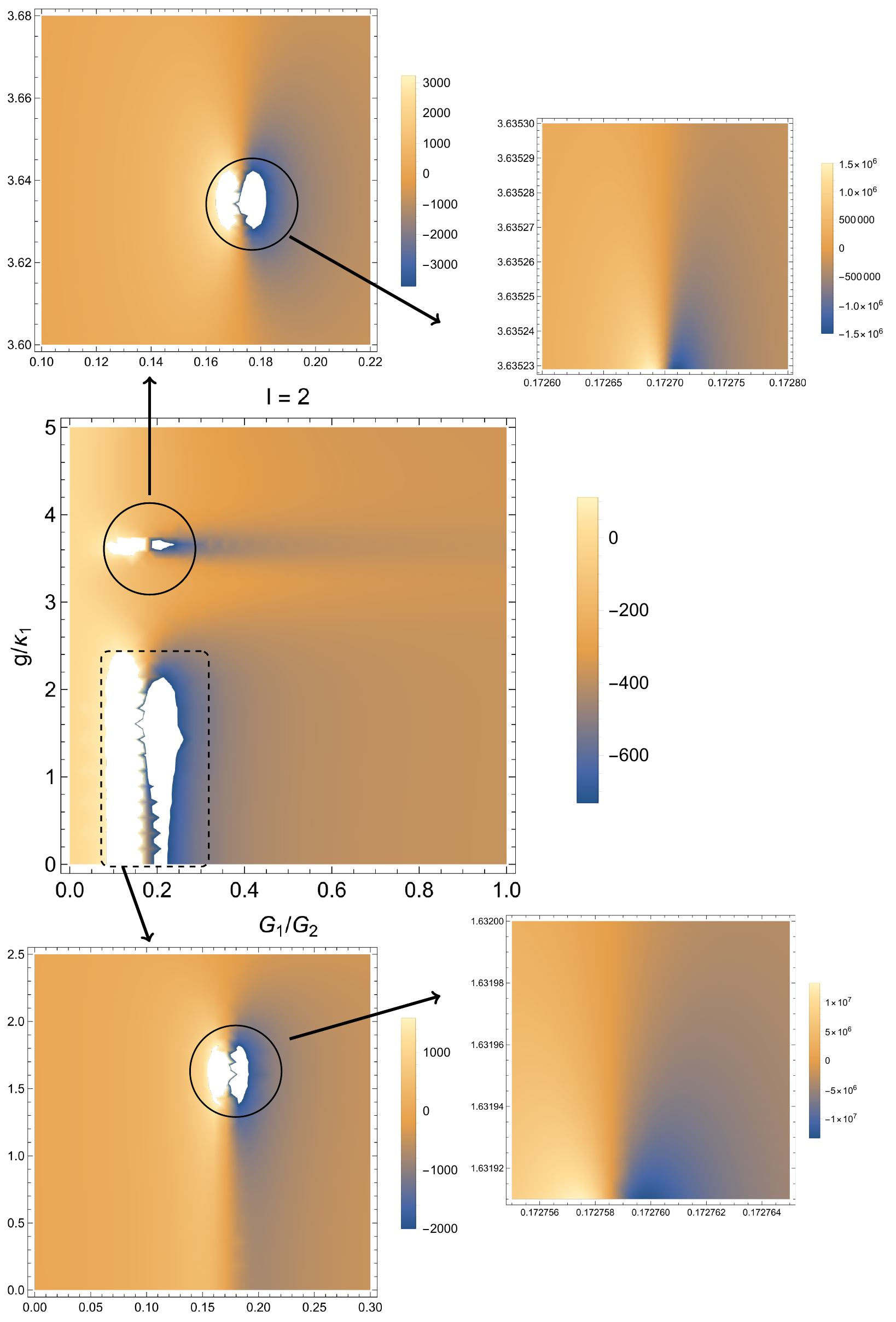}
	\caption{Illustrarion of group delay in Ising anyons when the first cavity is deformed by strength of  {$q=\textrm{exp}\left(\pi\textrm{i}/2\right)$}, at $\omega_{\text{m}1}=\omega_{\text{m}2}$, with normalized inter-cavity transmission strength (y-axis) and ration of optomechanical coupling strength of two cavities. The marked white regions in the density plot denotes region of extremely enhanced fast-slow light.}
	\label{fig:delay-ising}
\end{figure}
\begin{figure}[tbh!]
	\centering
	\includegraphics[width=0.6\linewidth]{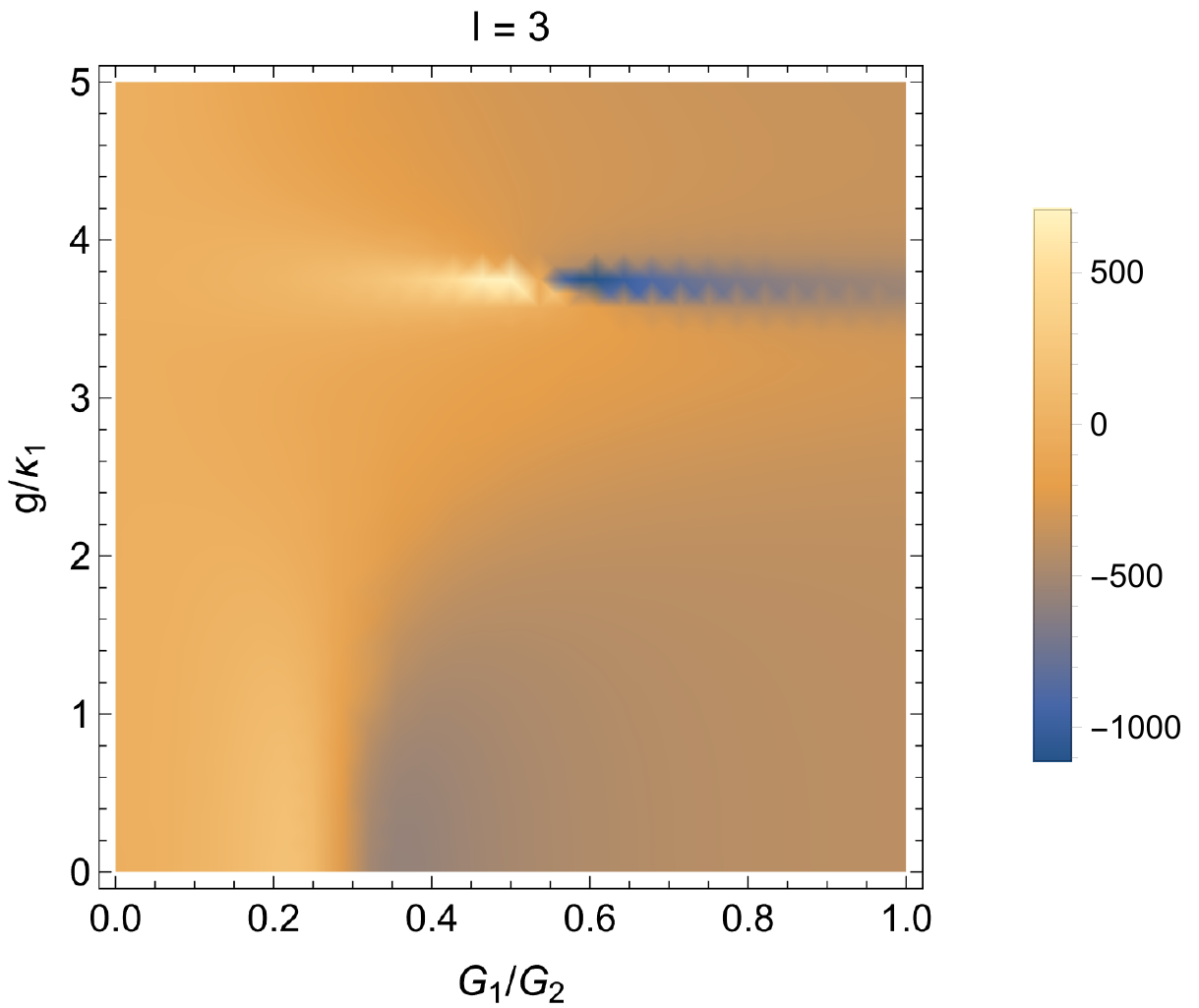}
	\caption{Illustrarion of group delay in Fibonacci anyons when the first cavity is deformed by strength of $ {q = \textrm{exp}\left(2\pi\textrm{i}/5\right)}$,
		 at $\omega_{\text{m}1}=\omega_{\text{m}2}$, with normalized inter-cavity transmission strength (y-axis) and ration of optomechanical coupling strength of two cavities. Although the fast and slow light is enhanced compared to non-deformed case but the enhancement can not go beyond the milliseconds region.}
	\label{fig:delay-fibonacci}
\end{figure}
\end{document}